\documentclass[twocolumn]{aastex631}

\hypersetup{
   colorlinks,
   linkcolor={blue!88!black!80},
   citecolor={blue!88!black!80},
   urlcolor={blue!88!black!80}}

\usepackage{color,subfigure}
\usepackage{float}
\usepackage{mathrsfs}
\usepackage{natbib}
\usepackage{xspace}
\usepackage{hyperref}
\usepackage{multirow}
\usepackage{booktabs}
\usepackage{float}

\usepackage{amsmath}

\newcommand{\RNum}[1]{\uppercase\expandafter{\romannumeral #1\relax}}

\newcommand{\hgamma}{H{$\gamma$}}
\newcommand{\hbeta}{H{$\beta$}}
\newcommand{\halpha}{H{$\alpha$}}

\def\FeII{Fe\,{\sc ii}}

\def\MgII{Mg\,{\sc ii}}
\def\HeII{He\,{\sc ii}}

\def \OIII {[O\,{\sc iii}]}

\newcommand{\OIIIb}{[O{\sevenrm\,III}]\,$\lambda$5007}

\newcommand{\SII}{[S{\sevenrm\,II}]}

   \font\sevenrm=cmr7 scaled 1000
\newcommand{\comments}[1]{}

\begin{document}

\title{Dormancy and Reawakening  Over Years: Eight New Recurrent Changing-Look AGNs }

\author[0000-0002-2052-6400]{Shu Wang}
\affiliation{Department of Physics \& Astronomy, Seoul National University, Seoul 08826, Republic of Korea; woo@astro.snu.ac.kr; wangshu100002@gmail.com}

\author[0000-0002-8055-5465]{Jong-Hak Woo}
\affiliation{Department of Physics \& Astronomy, Seoul National University, Seoul 08826, Republic of Korea; woo@astro.snu.ac.kr; wangshu100002@gmail.com}

\author[0000-0002-8055-5465]{Elena Gallo}
\affiliation{Department of Astronomy, University of Michigan, Ann Arbor, MI 48109, USA}

\author[0000-0002-8055-5465]{Donghoon Son}
\affiliation{Department of Physics \& Astronomy, Seoul National University, Seoul 08826, Republic of Korea; woo@astro.snu.ac.kr; wangshu100002@gmail.com}

\author[0000-0002-6893-3742]{Qian Yang}
\affiliation{Center for Astrophysics $|$ Harvard \& Smithsonian, 60 Garden St., Cambridge, MA 02138, USA}

\author[0000-0002-8402-3722
]{Junjie Jin}
\affiliation{Key Laboratory of Optical Astronomy, National Astronomical Observatories, Chinese Academy of Sciences, Beijing 100012, People's Republic of China}

\author[0000-0001-8416-7059]{Hengxiao Guo}
\affiliation{Key Laboratory for Research in Galaxies and Cosmology, Shanghai Astronomical Observatory, Chinese Academy of Sciences, 80 Nandan Road, Shanghai 200030, People's Republic of China}

\author{Minzhi Kong}
\affiliation{Department of Physics, Hebei Normal University, Shijiazhuang 050024, People's Republic of China}

\begin{abstract}

We report the discovery of eight new recurrent changing-look (CL) active galactic nuclei (AGNs), including seven re-brightening turn-off AGNs and one fading turn-on AGN. These systems are valuable for placing constraints on the duration of dim and bright states, which may be linked to the AGN duty cycle or disk instability. 
Long-term optical light curve analysis reveals that many objects in our sample exhibit a prolonged plateau during the dim states lasting 4 to 7 years, with gradual turn-on/off process. We observe no significant difference between the turn-on and turn-off timescales, and this timescale is broadly consistent with the heating/cooling front propagation timescale. The comparison between optical and infrared variations supports that these transitions are driven by changes in accretion disk emission rather than dust obscuration.  Our discovery significantly increases the previously identified recurrent CL AGN sample from eleven objects to nineteen, demonstrating that some AGNs can enter dormancy and reawaken on timescales of a few years, which provides useful information for understanding AGN episodic accretion.

\end{abstract}


\section{Introduction}\label{sec:intro}

Active Galactic Nuclei (AGNs) and quasars, which are fueled by accretion onto supermassive black holes, release tremendous energy across a wide range of wavelengths. In the optical band, the most prominent feature is the presence of strong emission lines ionized by the blue continuum from the accretion disk.  Based on the presence/absence of broad emission lines, AGNs are classified into two types, i.e., Type 1/2, respectively. This classification is traditionally attributed to orientational effects \citep{Antonucci93}, i.e.,  Type 2 AGNs are observed edge-on so the observer's line of sight is obscured by a dusty torus, while Type 1 AGNs are observed relatively face-on and free from torus obscuration. The Eddington ratio may also play an important role in determining the torus covering factor \citep[e.g.,][]{Lawrence91,Oh15,Ricci22b}, suggesting a radiation-regulated evolutionary sequence, more so than a Type 1 vs. 2 dichotomy.

In some special cases,  known as changing-look (CL) AGNs or changing-state AGNs \citep[e.g.,][]{Graham20,Ricci22}, AGNs appear to transition between Type 1 and 2 ``states", or vice versa.  While early CL AGNs were discovered serendipitously, the growing body of large-area spectroscopic surveys and time-domain photometric surveys has enabled systematic searches that significantly increased the sample size \citep[e.g.,][]{Macleod16,Macleod19,Runnoe16,Yang18,Yang24, Graham20,Sanchez21,Guo20,Green22, Hon22, Guo24,Guo24b,Zeltyn24,Dong24}. Many of these systems are selected via extreme continuum variability \citep[e.g.,][]{Macleod19}.
Long-term light curve characteristics have also proven effective in identifying CL AGNs \citep{Navas22,Navas23b,Wang24}, especially for host-dominated cases.

There is increasing consensus that the CL AGN phenomenon cannot be attributed to variations in dust obscuration \citep[e.g.,][]{Sheng17,Hutsemekers19,Yang23,Temple23}.  Most studies favor changes in the accretion rate as the most plausible explanation, although the observed transition timescales (with an upper limit of $10 \sim 20$ yrs) are considerably shorter compared to the viscous timescale of the thin disk model \citep[10$^{4\sim7}$ years;][]{Krolik99} which is required for major changes in the global accretion rate.  In fact, recent well-sampled observations have revealed that the transition timescale of CL AGNs can be as short as months to years \citep{Yan19,Zeltyn22}.  %Studying the spectral evolution and long-term variability will shed light on accretion disk physics.
To resolve this discrepancy, mechanisms such as thick disks or strong magnetic fields \citep[e.g.,][]{Jiang19,Dexter19,Li-Cao19,Feng21c} are proposed to shorten the radial transport timescale.

Among the CL AGN sample, some rarer cases have been observed to undergo a complete cycle of turning on and off, or vice versa \citep[see the eleven objects collected by][and references therein]{Wang-J24}, which are referred to as recurrent CL AGNs.  One of the best examples is Mrk 590 (see also Mrk 1018), which has been extensively monitored over the past 4 decades. Mrk 590 was classified as a Type 1.5 AGN $\sim$50 yrs ago \citep{Osterbrock77}, transitioned to Type 1.0 state in 1980s and remained in the bright state until mid 1990s. It then faded by a factor of 100, and appeared as a Type 2 AGN by 2013 \citep{Denney14}. In 2017, it returned to a Type 1 AGN \citep{Mathur18,Raimundo19}.

These recurrent CL AGNs are valuable for placing stringent constraints on the duration of AGN bright and dim states. If these transitions directly reflect significant changes in accretion rates, they offer important insights into the AGN duty cycle and episodic accretion. It is generally thought that AGN activity evolves over timescales of about $10^6$ years; however, the recurrent CL AGN sample, exemplified by Mrk 590, suggests that some AGNs can enter dormancy and return to activity on much shorter timescales. Alternatively, some recurrent CL AGNs,  such as those with quasi-periodic outbursts (e.g., NGC 1566), may be explained by the radiation pressure instability model \citep{Sniegowska20, Pan21}. Long-term light curves allow for the comparison of turn-on and turn-off timescales, providing valuable evidence to test these models.

In this work, we identify eight new recurrent CL AGNs, nearly doubling the previously known sample of recurrent CL AGNs. In \S\ref{sec:Sample}, we detail our sample selection process and follow-up spectroscopic observations. In \S\ref{sec:results}, we analyze the spectral evolution and long-term variability of these recurrent CL AGNs. We discuss the implications of our recurrent CL AGN sample in \S\ref{sec:discussion} and summarize in \S\ref{sec:summary}. Throughout this paper, we adopt a flat cosmological model with $h_0=70.0$ and $\Omega_m=0.3$.

\begin{table*}[htbp]
    \centering
    \caption{Summary of Sample Properties and Follow-up Spectroscopic Observations}
    \label{tab:information}
    \begin{tabular}{c c c c  c  c c c c}
     \hline \hline
    ID & SDSS Name  &  $z$ & $r$-band &  MJD  & Telescope/Instrument & $t_{\rm exp}$ & Transition History  & Reference \\ 
    & & & (mag) &    &  & (s)  &   &   \\ 
    (1) & (2) & (3) &  (4)  & (5) & (6)  & (7)  & (8) & (9) \\ 
    \hline
    
    ID01 & J002311.06+003517.5  & 0.4211 &  18.7 & 59826 & Gemini/GMOS & 900 & Turn-off  & 1 \\ 
    ID02 &J013458.36$-$091435.4  & 0.4430 & 18.9 &  59856 & MDM/VPH & 3600 & Turn-off & 2 \\
    %J084957.78+274728.9 & 0.2990 &  19.0 & 58846  & eBOSS/BOSS &  \nodata  & \citet{Yang18} \\
    ID03 &J101152.98+544206.4  & 0.2464 &  18.8  &  59913 & MDM/VPH & 3600 & Turn-off & 3\\
    ID04 &J110423.21+634305.3  & 0.1643 &  19.3  &  59986 & MDM/VPH & 3600 & Turn-off & 4\\
    ID05 &J110456.02+433409.2  & 0.0490 &  16.9 &  59868 & MDM/VPH & 3600 & Turn-on & 3 \\
    ID06 &J115039.32+362358.4  & 0.3400 &  19.6 &  59753 & Gemini/GMOS & 900 & Turn-off & 4\\ 
    ID07 &J144702.87+273746.7  & 0.2241 &  18.0 &  69715 & MDM/VPH & 3600 & Turn-off & 2\\
    ID08 &J161711.42+063833.5  & 0.2290 &  18.5 &  59738 & Gemini/GMOS & 900 &  Turn-off & 2\\
    \hline
   
\multicolumn{9}{p{17.5cm}}{Note. Column1: The object ID assigned in this work. Column 2: the SDSS name of each object. Column 3: redshift. Column 4: the median $r$-band magnitude of ZTF light curves. Column 5 and 6: the MJD and the telescopes of the follow-up spectroscopic observations. Column 7: the exposure time $t_{\rm exp}$ in the unit of seconds. Columns 8 and 9: the transition history and the references, respectively. 
References. 1: \citet{Green22}; 2: \citet{Macleod19}; 3. \citet{Runco16}; 4: \citet{Yang18}.}
 \end{tabular}
\end{table*}

\section{Sample and Follow-up observations}\label{sec:Sample} 

\subsection{Sample selection}\label{sec:method}

We start from the sample of known CL AGNs discovered before 2022, resulting in a total of nearly 200 systems.  To search for recurrent CL AGNs, we adopt the same methodology proposed by \citet{Wang24}, which compares the historical spectral types with the current light curve characteristics.  We  examine the variability behavior of known CL AGNs, utilizing $g$-band light curves from the public data release (DR) 15 of Zwicky Transient Facility \citep[ZTF;][]{Bellm19}. The $\sigma_{\rm QSO}$ parameter \citep{Butler11} is calculated to categorize their light curves, where sources with large $\sigma_{\rm QSO}$, e.g., $\sigma_{\rm QSO}>3.0$,  indicate typical Type 1 AGN light curve (see \citet{Wang24} for more details).  We find that, within the known turn-off AGNs, about 60\% sources exhibit $\sigma_{\rm QSO}>3.0$ based on their ZTF light curves, suggesting that they potentially already transitioned back to bright states. On the other hand, some turn-off sources display flat and scattered ZTF light curves with $\sigma_{\rm QSO}<3.0$, suggesting that they remain in dim states during the ZTF observing period. For turn-on AGNs, the majority of objects still retain Type 1 light curve characteristics, while some rare cases exhibit flattened light curves, indicating that they have returned to a dim state. In our pilot survey, we performed follow-up spectroscopic observations for seven re-brightening turn-off sources and one fading turn-on source. The information and the references of these sources are summarized in Table \ref{tab:information}.

\subsection{Follow-up Spectroscopic Observations}\label{sec:observation}

Follow-up spectroscopic observations were conducted using the 2.4m Hiltner telescope at the MDM Observatory in Kitt Peak, Arizona, USA, and the Gemini-North 8m telescope at Mauna Kea, Hawaii, USA. 

The MDM observations utilized the Volume Phase Holographic (VPH) red grism and a 4$^{\prime\prime}$ slit in the outer position, providing spectral coverage from 4000 to 9000$\,$ \AA\ with a spectral resolution of 370. On-source exposure times ranged from 2700 to 3600 seconds, depending on the target’s brightness, yielding a typical signal-to-noise ratio (SNR) of 15 per pixel. The Gemini-North observations employed the GMOS spectrograph with a configuration consisting of a 2$^{\prime\prime}$ slit, an R400 disperser, and a CG455 blocking filter to minimize contamination from higher-order spectra. A two-pixel binning mode was used in both spatial and wavelength directions, providing spectral coverage from 5000 to 9500$\,$\AA\ and a spectral resolution of $R \sim 480$. Exposure times were set at approximately 900 seconds for $r$-band targets of $\sim$19.0 magnitude, achieving a typical SNR of 60 per pixel.

For historical spectra, we compiled available data from SDSS DR 18 \citep{Almeida23} and the LAMOST survey \citep{Zhao12,Luo15} DR 9. Necessary corrections to the connection between the blue and red-arm LAMOST spectra were applied based on the experience from previous works \citep[e.g.,][]{Ai16,Dong18,Yao19,Jin23}. For sources identified using other facilities, we digitized their spectra from the original literature using the software {\tt WebPlotDigitizer}\footnote{\url{https://automeris.io}} with our best efforts.

To ensure consistent flux levels, the spectra from various epochs were rescaled using the \OIII\ flux, which is expected to remain constant over timescales of 10 to 20 years, as narrow-line regions typically span kiloparsec scales \citep{Bennert02}. The SDSS spectra were adopted as the \OIII\ flux reference, and spectra from other facilities were rescaled when the flux difference exceeded 10\%.

\section{Results }\label{sec:results} 

\begin{figure*}[h!]
    \centering
    \includegraphics[width=0.99\linewidth]{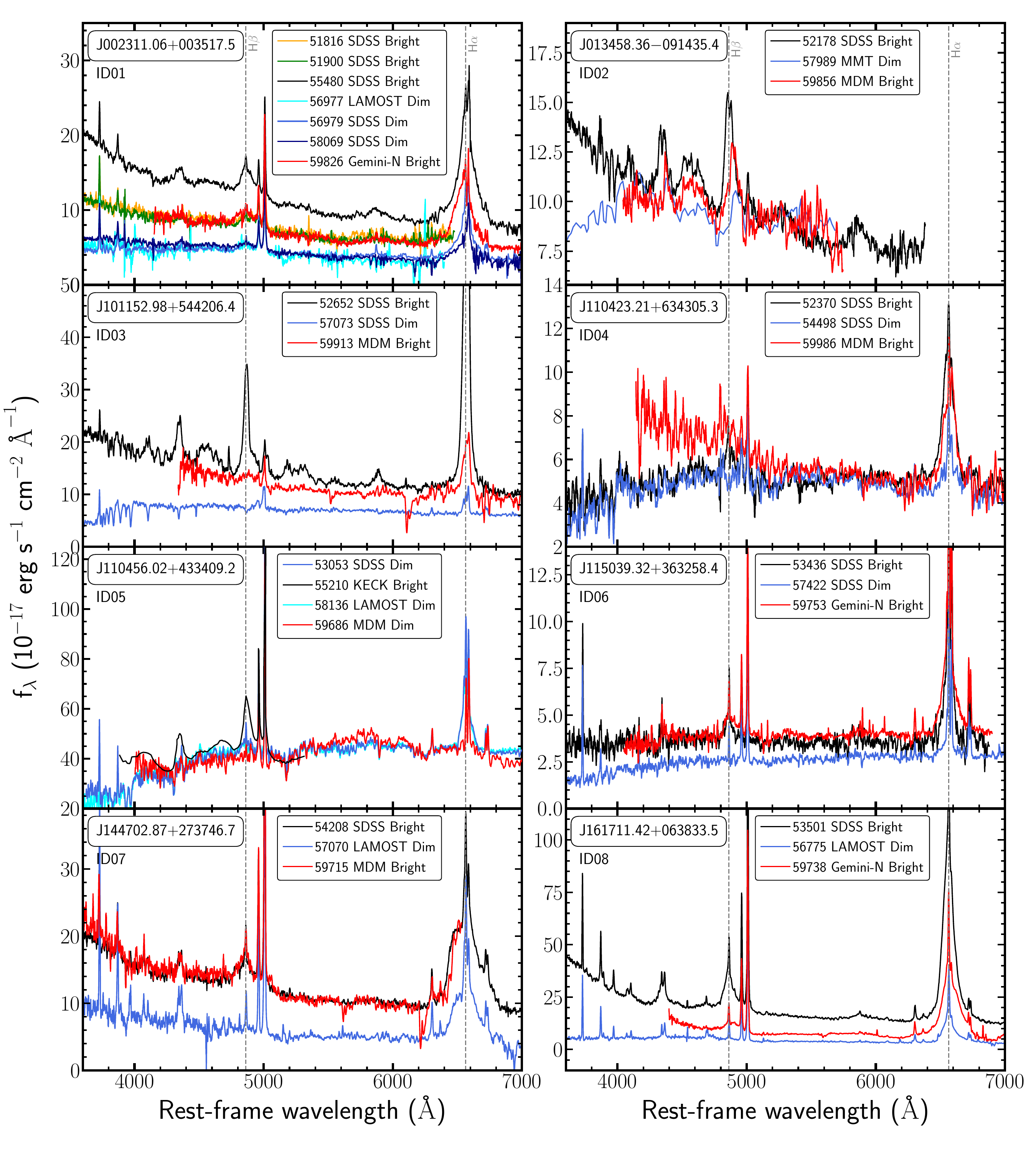}

    \caption{The 5-pixel smoothed spectra for the eight recurrent CL AGNs studied in this work. The SDSS name of each target is labeled on the top left corner along with the object ID assigned in this work. Spectra from different epochs are presented using different colors, with the corresponding epoch and telescope information provided in the upper middle legend.  The rest-frame wavelengths of \hbeta\ and \halpha\ are labeled using vertical dashed lines.}
    \label{fig:spectra}
\end{figure*}

\begin{figure*}[htbp]
    \centering
    \includegraphics[width=0.99\linewidth]{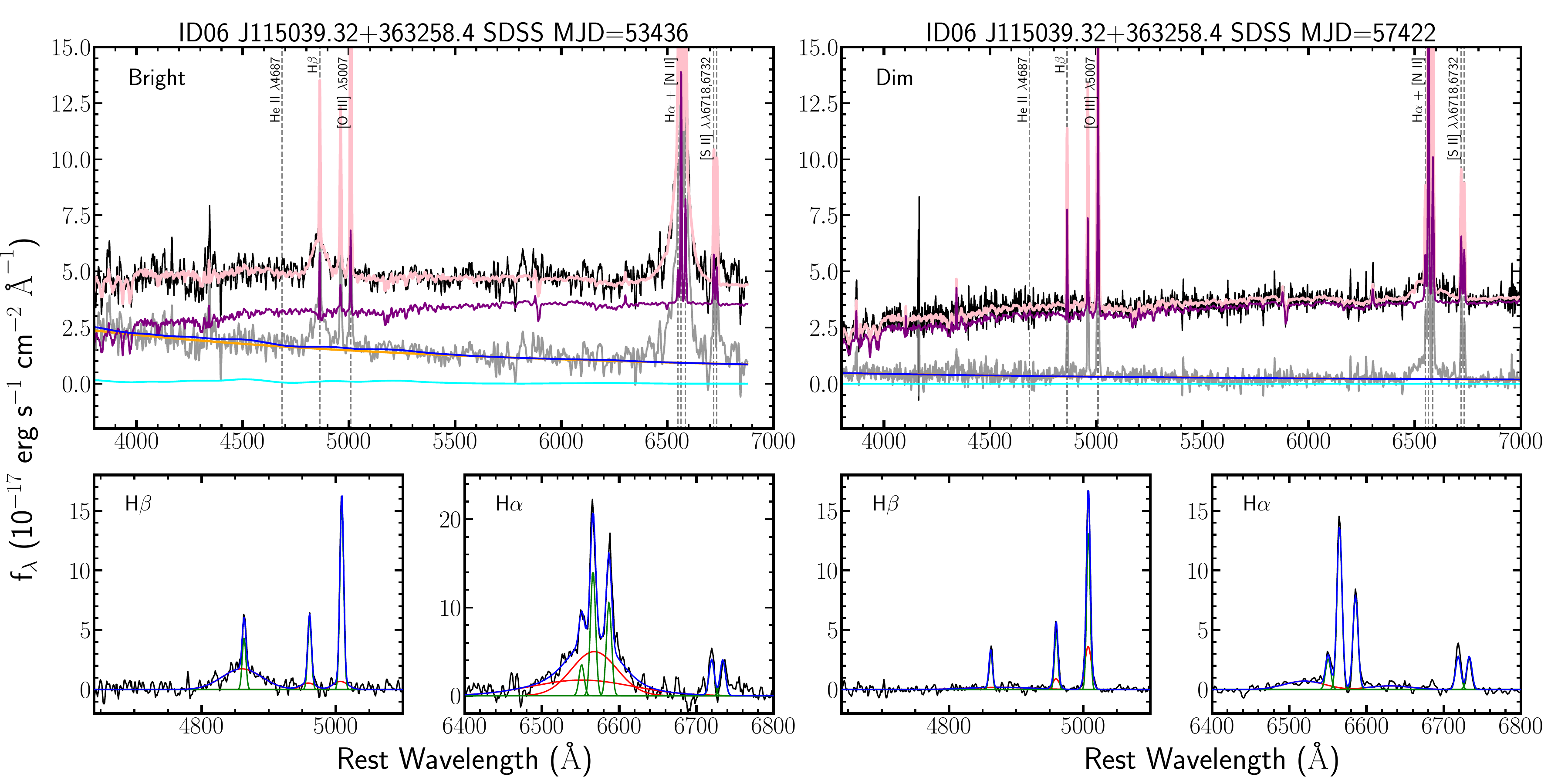}
    
    \caption{An example of our spectral decomposition  based on {\tt PyQSOFit} for the bright (left) and dim (right) state spectra of SDSS J115039.32+363258.4. In the upper panels,  the original spectra (black) are decomposed into host galaxy (magenta) and AGN  (grey) component using prior-informed PCA methods.  The AGN continuum model consists of a power-law component (orange) and \FeII\ template (cyan), and the total continuum model is shown in blue. In the lower  panels, we show the line-fitting results for the  \hbeta\ (left) and \halpha\ (right) line complex. The red and green lines represent the broad and narrow Gaussians, while the blue lines represent the sum of these models. }
    \label{fig:decomposition}
\end{figure*}

\begin{figure*}[htbp]
    \centering
    \includegraphics[width=0.44\linewidth]{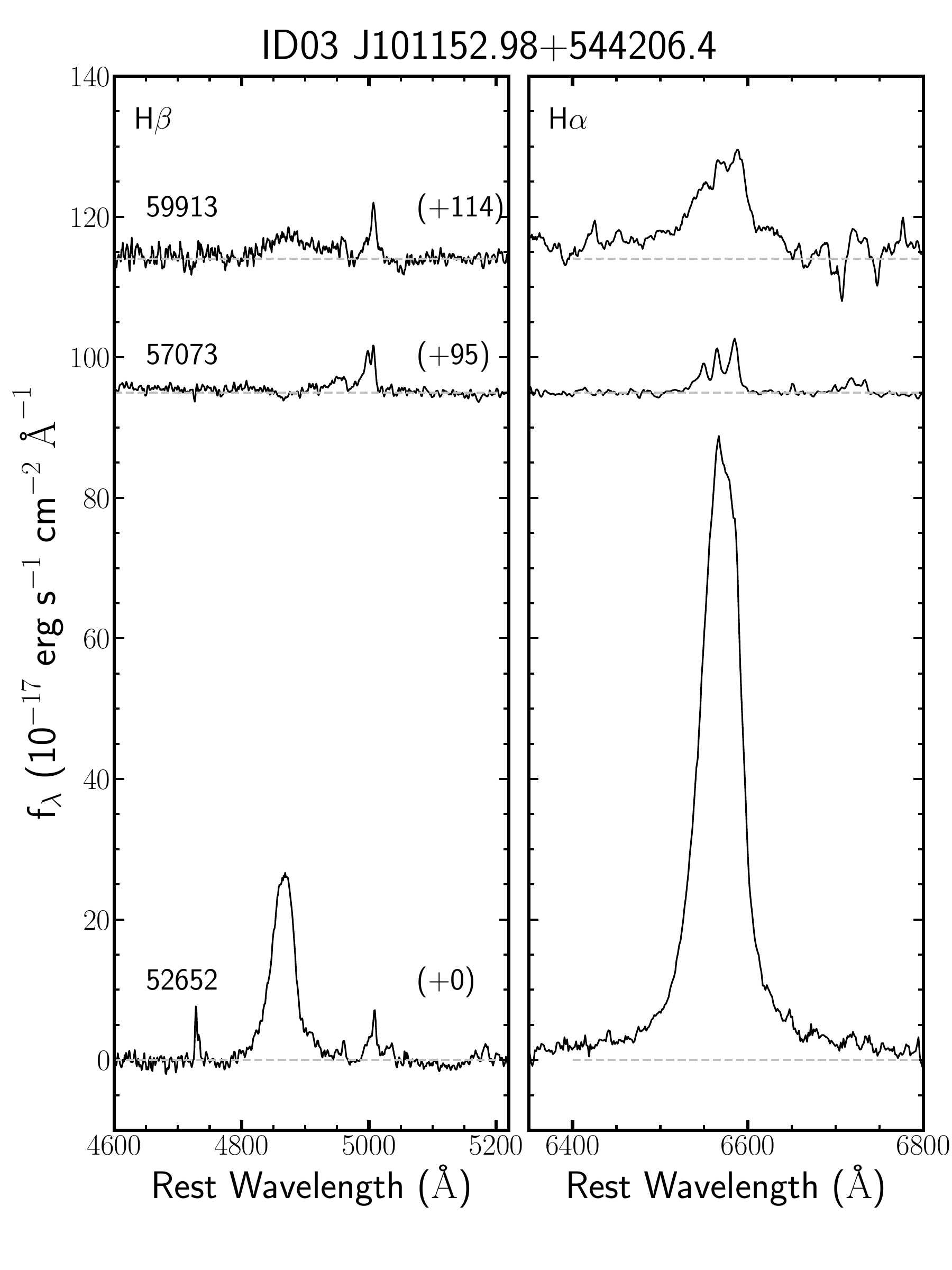}
    \includegraphics[width=0.44\linewidth]{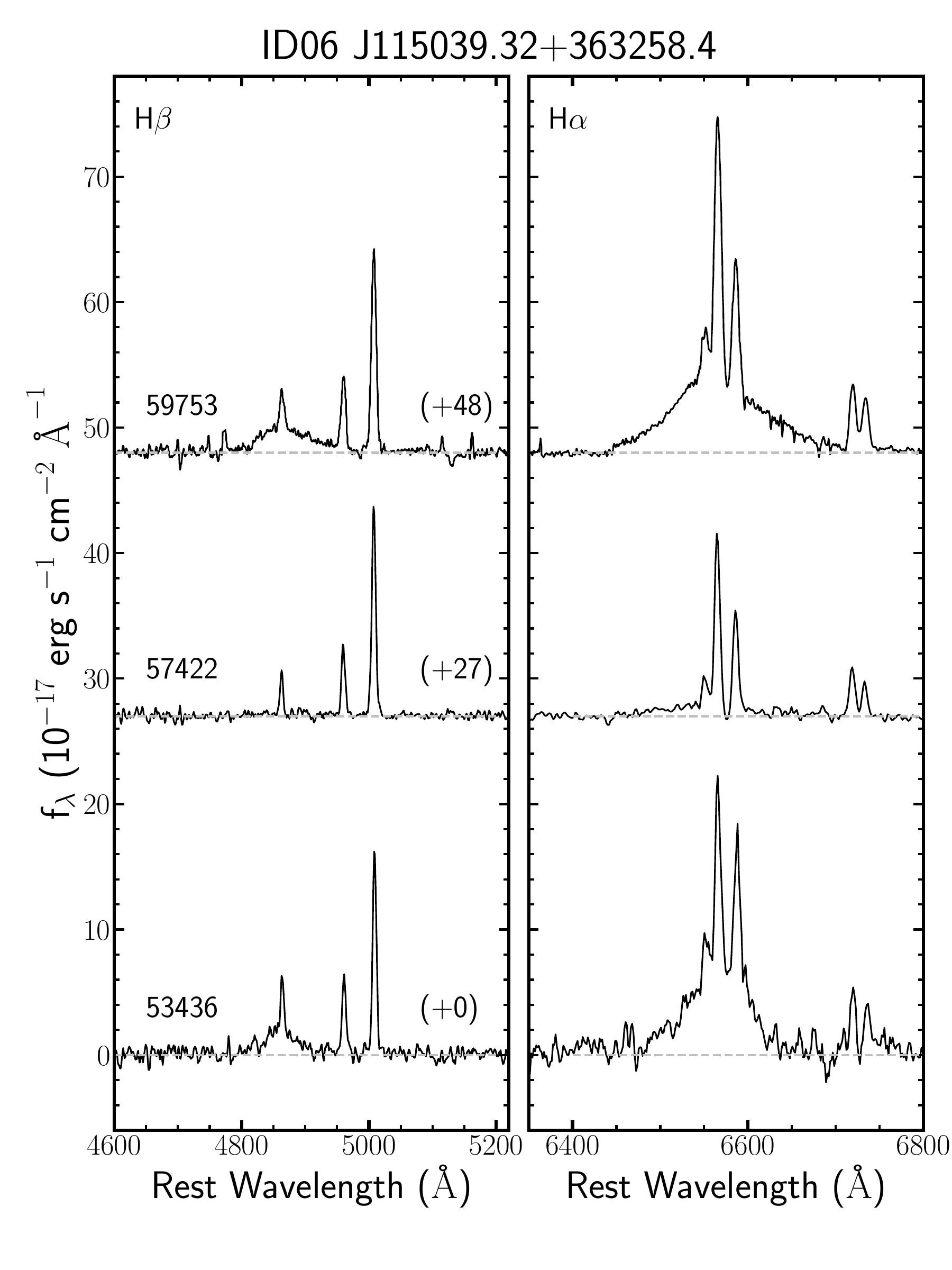}

    \caption{Two examples of spectral evolution of \hbeta\ (left panels) and \halpha\ (right panels). The left object is J101152.98+544206.4 while the right object represents J115039.32+363258.4. The MJD of each spectrum is labeled and the spectra are shifted vertically (by the amount in the brackets) for display purposes. The gray dashed lines represent level of zero flux in the shifted line-only spectrum, where the host galaxy and AGN continuum model have been subtracted. }
    \label{fig:lineprofile}
\end{figure*}

\subsection{Spectral Evolution}\label{sec:Spectral_evolution}

%\section{Spectral decomposition}

Figure \ref{fig:spectra} presents a comparison of multi-epoch spectra for the eight recurrent CL AGNs in our sample. Significant variations in the \hbeta\ and \halpha\ line strengths are observed. All seven re-brightening turn-off CL AGNs exhibit the disappearance and subsequent reemergence of broad \hbeta\ profiles, indicating the capture of a full cycle from turn-off to turn-on transition. The broad \halpha\ profiles also show significant variations and consistent trends; however, in some cases, they remain detectable even during the dim states. In general, we find that \hbeta\ tends to disappear more readily in dim-state spectra compared to \halpha. This may be because weaker lines are more likely to fall below the detection limit (e.g., due to SNR), or it could be related to ionization evolution, as discussed by \citet{Guo19, Guo20a} using locally optimally emitting clouds simulations. For the only fading turn-on CL AGN in our sample, the most recent spectrum shows an almost complete disappearance of both \hbeta\ and \halpha, indicating a turn-off transition after the turn-on event. %In general, we find that \hbeta\ tends to disappear more readily in dim-state spectra compared to \halpha. This may be because weaker lines are more likely to fall below the detection limit (e.g., due to SNR), or it could be related to ionization evolution, as discussed by \citet{Guo19, Guo20a} using locally optimally emitting clouds simulations.

We performed spectral decompositions for each spectrum using the PyQSOFit package \citep{Guo18,Shen19}. We adopted the latest version of PyQSOFit \citep{Ren24} which incorporates prior-informed Principal Component Analysis  (PCA) for AGN-host decomposition, enabling more accurate subtraction of the host galaxy. We adopted the first 10 galaxy components and the first 10 quasar components created by \citet{Yip04a,Yip04b} in PCA. Figure \ref{fig:decomposition} shows an example of the decomposition result for both the bright and dim-state spectra.  After subtracting the host component, we fit the residual spectrum with a power law plus an \FeII\ template \citep{Boroson92} within line-free windows \citep{Shen11,Wang20}. Following this, we subtracted the best-fit AGN continuum model and performed line fitting for \hbeta\ and \halpha\ line complex separately. A single Gaussian was used for each narrow line, while two Gaussians were adopted for each broad line to account for any asymmetry. The scale and width of narrow \hbeta\ and \halpha\ are tied to adjacent strong narrow emission lines, e.g., \OIIIb, and \SII, respectively. 

We measured the AGN continuum luminosity at 5100 \AA\ ($L_{5100}$) and the full-width-at-half-maximum (FWHM) of \hbeta\  and \halpha\ if detected. Except for two objects (J002311.06+003517.5 and J013457.36$-$091435.4), the majority of our sample show $L_{5100}<10^{44}$ erg s$^{-1}$, placing them in the Seyfert regime. The BH mass is measured using the BH mass estimator based on the bright-state $L_{5100}$ and \hbeta\  FWHM \citep{Shen12}.  The  derived black hole masses range from $10^{7.2}$ to $10^{8.8}$ $M_{\odot}$. The bolometric luminosity is derived from $L_{5100}$ using a constant correction factor of 9.26 \citep{Shen11}. The Eddington ratios are then calculated by dividing the bolometric luminosity by the Eddington luminosity. The median Eddington ratio of our sample are $0.04$ and $0.003$ for bright and dim states, respectively. 

Figure \ref{fig:lineprofile} presents two examples of the \hbeta\ and \halpha\ profile evolution, clearly illustrating the significant time evolution of line strength within our sample. 

\subsubsection{Description for Each Object}

\textit{J002311.06+003517.5 (ID01)}: This object was identified as a turn-off CL AGN by \citet{Green22}. Early SDSS spectra from 2002 (MJD = 51816 and 51900) displayed a clear broad \hbeta\ component with moderate line strength, which we classify as bright states. The \halpha\ line was not covered in these observations. At MJD = 55480, the \hbeta\ line became approximately 1.6 times stronger, and a strong broad \halpha\ profile was observed. Around MJD = 56978, the object transitioned to a dim state where the broad \hbeta\ and \halpha\  became significantly weaker, although still present. Notably, the broad \halpha\ exhibited a pronounced asymmetric profile. \citet{Green22} reported two additional epochs (MJD = 57597 and 58037, observed with Magellan and MMT, respectively) that are not shown in Figure \ref{fig:spectra}. During these epochs, the broad \hbeta\ nearly vanished, indicating a transition to a Type 1.9 AGN. In our follow-up spectrum taken at MJD = 59826, the object returned to a bright state with moderately strong \hbeta\ and \halpha\ lines, resembling the early SDSS spectra in 2002.

\textit{J013458.36$-$091435.4 (ID02)}: This object was identified as a turn-off CL AGN by \citet{Macleod19}. The early SDSS spectra from 2002 (MJD=52178) revealed a strong broad \hbeta\ component, which became substantially weaker by MJD=57989 \citep{Macleod19}. Although the broad \hbeta\ remained detectable in its dim state, \citet{Macleod19} noted that the variability SNR of \hbeta\ flux was above 3, indicating a significant change in line flux. In our follow-up spectra taken at MJD=59856, the broad \hbeta\ component was considerably stronger, although still not as prominent as in the early SDSS spectra.  Notably,  the flux decrease (or enhancement) is coupled with a corresponding increase (or decrease) in asymmetry, suggesting this object may represent a special case among CL AGNs.

\textit{J101152.98+544206.4 (ID03)}: This object was identified as a turn-off CL AGN by \citet{Runnoe16}. It shows dramatic changes in the continuum and emission line flux, where the continuum luminosity at 5100 \AA\ dropped by a factor of $\sim$10 within a time span of 10 years \citep{Runnoe16}. In the dim state (MJD=57073), the  broad \hbeta\ completely disappeared below the detection limit and the broad \halpha\ was also super weak, which dropped by a factor of $\sim50$ compared to the bright state. In our follow-up spectra taken at MJD=59753, we observed the reemergence of both broad \hbeta\ and \halpha, suggesting this AGN reawakened in $\sim10$ years. The broad \halpha\ flux increased by a factor of 8 compared to the dim-state spectrum, but are still approximately a factor of 6 weaker compared to the earlier bright state spectrum. This may suggest that it  takes time for broad-line region (BLR) to grow after reawakening from dormancy. 

In addition, it's noteworthy that the \OIIIb\  displayed a double-peaked profile in the dim state. Consequently,  the FWHM of \OIIIb\  in the dim state (1180$\pm89$ km s$^{-1}$) is much larger than those in the two bright states ($\sim$700$\pm35$ km s$^{-1}$).  These differences are likely resulted from  an enhancement of the blue wing \OIII\  component during the dim state, possibly linked to outflows associated with the disk state transition, e.g., the  transition into an ADAF.

\textit{J110423.21+634305.3 (ID04)}: This object was identified as a turn-off CL AGN by \citet{Yang18}. In the early SDSS spectrum (MJD = 52370), it displayed strong broad \hbeta\ and \halpha. At MJD=54498,  it transitioned to a dim state, where no broad \hbeta\ was detected and the broad \halpha\ was much weaker. In our follow-up spectra taken at MJD=59968,  both broad \hbeta\ and \halpha\ were clearly detected, suggesting a return to bright state.

\textit{J110456.02+433409.2 (ID05)}: this object was identified as a turn-on CL AGN by \citet{Runco16}, although it primarily underwent transitions between intermediate types, such as  shifting from Type 1.5/1.8 to Type 1.0 \citep{Runco16}. Compared to the early SDSS spectrum (MJD = 53436), the Keck spectrum obtained at MJD = 55210 exhibited a significantly stronger broad \hbeta\ line, as well as more pronounced \hgamma\ and \HeII\ lines. At MJD=58136, the LAMOST spectrum returned to a dim (medium) state similar to the earlier SDSS one. In the most recent spectra taken at MJD=59686 by MDM, both the broad \hbeta\ and \halpha\ lines almost completely disappeared, indicating a transition into a Type 2 AGN.

\textit{J115039.32+362358.4 (ID06)}: This object was identified as a turn-off CL AGN by \citet{Yang18}. In the early SDSS spectrum (MJD = 52370), both broad \hbeta\ and \halpha\ were clearly detected. At MJD = 54498, the broad \hbeta\  completely vanished, and the broad \halpha\ was a factor of 11 weaker, indicating a transition to a dim state. In our follow-up spectra taken at MJD=59968, both broad \hbeta\ and \halpha\ reappeared, with slightly stronger strength and broader line widths than in the earlier bright state spectrum. 

\textit{J144702.87+273746.7 (ID07)}: This object was identified as a turn-off CL AGN by \citet{Macleod19}.  In the early SDSS spectrum (MJD = 54208), it showed a typical Type 1 spectrum with strong broad \hbeta\ and \halpha. According to the LAMOST spectrum taken at MJD=54498,  it transitioned to a dim state (Type 1.9), where no broad \hbeta\ was detected, while broad \halpha\ was still quite strong. Additional spectrum taken by Polomar in 2017 \citep{Macleod19} showed consistent trends with the LAMOST spectrum. In our follow-up spectra taken at MJD=59715, the broad \hbeta\ returned and the spectrum resembles the early bright-state spectrum.  

\textit{J161711.42+063833.5 (ID08)}: This object was identified as a turn-off CL AGN by \citet{Macleod19}. In the early SDSS spectrum (MJD = 53501), it showed strong broad \hbeta\ and \halpha. At MJD = 56775, the LAMOST spectrum showed no broad \hbeta\ and the broad \halpha\ was a factor 7 weaker, suggesting a transition to a Type 1.9 AGN.  The  spectrum taken in 2016 by Magellan \citep{Macleod19} are consistent with the LAMOST spectrum. In our follow-up spectra taken at MJD=56775, we observed reemergence of broad \hbeta\  and a factor of 3  stronger broad \halpha.

\subsection{Long-term Photometric Variability} \label{sec:lcs}

\begin{figure*}
    \centering
    \includegraphics[width=0.97\linewidth]{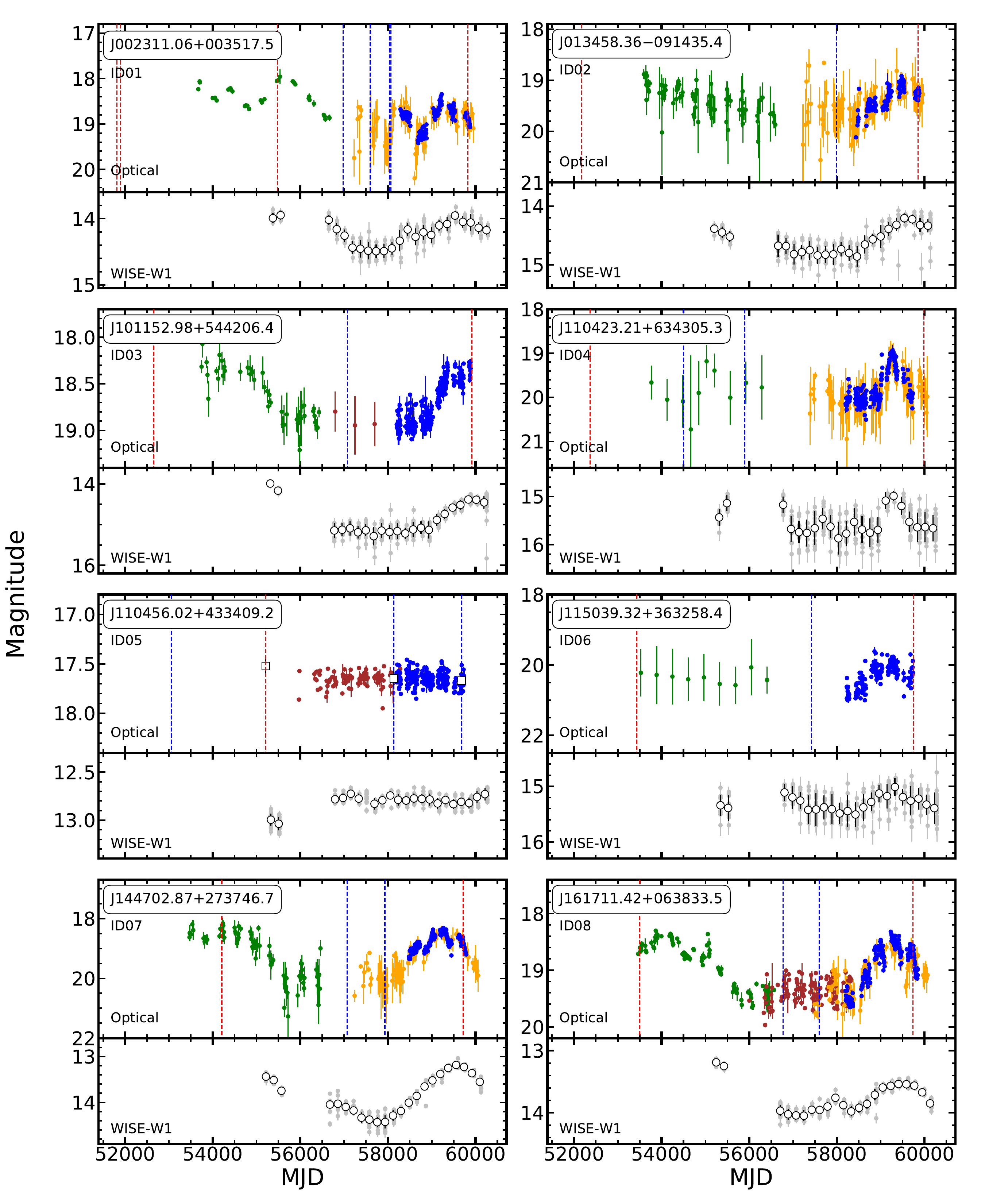}
    
    \caption{The long-term optical  (upper subpanels) and infrared (lower subpanels) light curves. The optical light curves are intercalibrated from CRTS (green), ASASSN (brown), ATLAS (orange), and ZTF (blue). The infrared light curves are based on WISE W1 band, with the open circles represent the median magnitude in every six months. The SDSS name of each target is labeled on the top left corner along with the  object ID assigned in this work. The vertical dashed lines indicate the spectroscopic observations, where red ones represent bright states while blue ones represent dim states. For J110456.02+433409.2, we overplot the synthetic magnitudes using open squares that are obtained from the recalibrated spectra. %The silver light curves are NEOWISE W1-band light curves that are manually shifted for displaying purposes.
    }
    \label{fig:LC}
\end{figure*}

In this section, we investigate the long-term variation of our sample. We collect the optical light curves from the Catalina Real Time Transient Survey \citep[CRTS;][]{Drake09}, the All-Sky Automated Survey for Supernovae \citep[ASASSN;][]{Kochanek17}, the Asteroid Terrestrial-Impact Last Alert System \citep[ATLAS;][]{Tonry18}, and ZTF. These light curves are then intercalibrated and merged using the software {\tt PyCALI} \citep{Li14}. 

Figure \ref{fig:LC} displays the merged long-term optical light curves. As previously mentioned, seven out of the eight objects are re-brightening turn-off AGNs and one object is fading turn-on AGN. For objects whose light curves reach the survey's limiting magnitude, we binned the light curves using 180 or 360 days width.

First, several re-brightening objects in our sample exhibit a clear long-duration platform during the dim states. For  example, J101152.98+544206.4 gradually transitioned to dim state at the end of CRTS light curve, stayed in dim states for 6.5 years, then smoothly returned to bright state during the ZTF period. 
Other examples include JJ144702.87+273746.7, J161711.42+063833.5, J110423.21+634305.3, and perhaps 013458.36$-$091435.4. The duration of the dim-state platform ranges from $4\sim7$  years in the rest frame. 
For the object (J002311.06+003517.5), significant flux fluctuations are still observed near its dim states, coinciding with dips in the light curve. This suggests the object may oscillate between bright and dim states over relatively short timescales, such as within months, or it may not have fully transitioned into a distinct state.  

J110456.02+433409.2 is the only fading object in our sample. Both its ASASSN and ZTF light curves are flat and scattered, suggesting that it potentially  transitioned to dim state at least since the beginning of ASASSN. %Thus,  the duration of the bright state is  (shorter than $\sim$9 years) is similar or shorter than the duration of the dim state.  

Second, we investigate the transition timescale for our sample.  Most of these objects  show a relatively slow and gradual turn-on/off process. The timescale for a one-magnitude variation (where available) ranges from approximately 1 to 9 years, with a typical value of 3–4 years. In addition,  there is no significant difference between the rising and fading timescales for individual objects. For instance, J144702.87+273746.7 took approximately 1500 days to fade by one magnitude, while it took roughly 1200 days to re-brighten by one magnitude.  Among our sample, J110423.21+634305.3 shows the most rapid variability with more than one magnitude variation within 1 year, which is more similar to a flaring event. It continued to show some variability in the following years and returned to flat and scattered light curves in 2024, as seen in the most recent ZTF DR 22 light curve. The subsequent light curve is different from that of a typical tidal disruption event.

\subsection{Optical and Infrared Variation}
One of the best ways to shed light on the mechanism of CL phenomenon is to compare the optical and infrared variations. A consistent trend between the optical and infrared light curves supports the variation in the accretion disk emission, while an anti-correlation  suggests changes in dust obscuration.  

In Figure \ref{fig:LC}, we compare the infrared and optical light curves  of our sample. For the infrared light curves, we use the W1-band data combining the ALLWISE and NEOWISE datasets. Our analysis reveals that the majority of our sample exhibit consistent trends between optical and infrared variations, i.e.,  lower infrared fluxes with lower optical fluxes and vice versa. This observation aligns with previous studies of CL AGNs and disfavors the changing obscuration scenario.  An exception is observed, J110456.02+433409.2, where the infrared flux increases as the AGN turns off.  However, this object has a close companion in projection which contaminates the NEOWISE photometry. We therefore consider this behavior as spurious.  In summary, our sample supports that the recurrent CL AGN phenomenon is related to the changes in accretion disk emission rather than dust obscuration.

\section{Discussion}\label{sec:discussion}

While it is clear that most CL AGNs are related to the changes in the accretion disk, the specific mechanisms causing the rapid and significant flux variation are still under debate.  Some insights can be obtained from the timescale arguments. In the thin disk model \citep{Shakura73}, the inflow timescale is given by
%\begin{equation}
\begin{align}
    t_{\rm inflow} \simeq 5000\, {\rm yr}\,  \left( \frac{\alpha}{0.02}\right)^{-1}   \left( \frac{\kappa}{\kappa_T}\right)^{-2}  \left( \frac{M_{\rm BH}}{10^8 M_{\odot}}\right)  \left( \frac{\dot{m}}{0.1}\right)^{-1}  \nonumber  \\
     \left( \frac{R}{100 r_g}\right)^{7/2} \label{equ:tinflow}
\end{align}
%\end{equation}
where $\alpha$ is the dimensionless viscous parameter, $\kappa$ is the opacity,  $\kappa_T$ is the value of Thomson scattering, $\dot{m}$ is the Eddington-scaled accretion rate,  and $r_g$ is the gravitational radius.  For the BH mass and accretion rates of our sample, the $t_{\rm inflow}$ is $10^{4}$ years at $R=100r_g$ (i.e., the optical emitting region).  However, the observed transition timescale and the duration of the dim state is on the order of several years.  This observed timescale is far too short for fluctuations in the accretion rate to propagate from the outer radius to the inner disk in the thin disk model \citep{laMassa15,Stern18,Graham20}. 

Previous studies have demonstrated that the observed timescale might be  consistent with  the propagation timescale  of heating/cooling front ($t_{\rm front}$) \citep{Stern18,Ross18,Graham20}. The $t_{\rm front}$  is parameterized as:
%\begin{equation}
%    t_{\rm th} = 1\, {\rm yr}\, \left( \frac{\alpha}{0.03}\right)^{-1} \left( \frac{M_{\rm BH}}{10^8 M_{\odot}}\right) \left( \frac{R}{150 r_g}\right)
%\end{equation}
\begin{equation}
    t_{\rm front} \simeq 11\, {\rm yr}\, \left( \frac{h/R}{0.05}\right)^{-1} \left( \frac{\alpha}{0.03}\right)^{-1} \left( \frac{M_{\rm BH}}{10^8 M_{\odot}}\right) \left( \frac{R}{100 r_g}\right)^{3/2}
\end{equation}
%\begin{equation}
%    t_{\rm vis} = 400\, {\rm yr}\,  \left( \frac{h/R}{0.05}\right)^{-2}\left( \frac{\alpha}{0.03}\right)^{-1} \left( \frac{M_{\rm BH}}{10^8 %M_{\odot}}\right) \left( \frac{R}{150 r_g}\right)^{3/2}
%\end{equation}
Using an average black hole mass of $10^{8.0}$ $M_{\odot}$ for our sample and default values for other parameters, we estimate $t_{\rm front}$ to be approximately 11 years,  which is within the same order of magnitude as our observations. Alternatively, the accretion disk may not be thin but instead thick, requiring a much shorter inflow timescale. The thick disk could  result from magnetic pressure \citep{Dexter19} or enhanced iron opacity \citep{Jiang16, Jiang20}. Additionally, magnetic field-driven disk outflows could also contribute to the shortening of the inflow timescale \citep[e.g.,][]{Feng21c}.

Several alternative mechanisms have been proposed for explaining the CL phenomenon, including the hydrogen ionization instability \citep{Noda18}, the disk state transition \citep{Noda18,Ruan19,Veronese24}, the radiation pressure instability between the inner ADAF and outer thin disk \citep{Sniegowska20,Pan21}, and the magnetic torque instability near the  innermost stable circular orbits \citep{Ross18}.  Distinguishing between these models is challenging, but comparing the turn-on and turn-off timescales may offer some clues.  In \S\ref{sec:lcs}, we observed that  the turn-on and turn-off timescales are not significantly different (see e.g., J144702.87+273746.7). However,  the thin disk model predicts a factor of 10 difference for our sample between the turn-on and turn-off  timescale,  as $t_{\rm inflow}$ is a function of $\dot{m}$ (see Equation \ref{equ:tinflow}). Similarly, some disk instability models such as the hydrogen ionization instability model predicts a rapid rise followed by an exponential decay  \citep[e.g.,][]{Dubus01},  which are different from our observations.  In general, the long-term optical light curves of our sample do not show significant differences from those of extreme variability AGNs. It is possible that the observed behavior of these CL AGNs represents the high-amplitude end of normal AGN variability \citep{Rumbaugh18}.
 
Regardless the mechanism responsible for the CL AGNs,   our work demonstrates that some AGNs can enter dormancy and return to activity on timescales of several years. AGNs are known to vary over a range of timescales and different evidence provides different indications.  Studies of extended emission regions suggest that the AGN activity varies on the timescale of $10^{6\sim7}$ years \citep[e.g.,][]{Lintott09}.  Based on the fraction of turned-off quasars \citep{Yang20b} and/or the relative frequency of broad \MgII\ emitters \citep{Roig14,Guo20a},  \citet{Shen21} estimated the average episodic quasar lifetime of hundreds of thousands of years.  Our finding suggests that the activity for some AGNs is not continuous even on shorter timescales. It is possible that these AGNs are in an unstable accretion phase, leading to extreme variability and occasional turning-off then rebrightening events, on timescales much shorter than predicted by steady accretion models.

Lastly, we note that there could be significant selection effects in our sample. On one hand, turn-off AGNs that have remained consistently in a dim state were not selected for observation. On the other hand, turn-off AGNs displaying persistent Type 1 variability were not favored in our selection, because some of them could be misidentified and never fully transitioned into a distinct state. A uniform investigation of known CL AGNs is essential for a comprehensive understanding of their long-term variations.

\section{Summary}\label{sec:summary}

In this work, we identified eight new recurrent CL AGNs, which consists of seven re-brightening turn-off AGNs and one fading turn-on AGN.
The follow-up spectroscopic observations were conducted using the MDM and Gemini telescope.  By analyzing the long-term optical light curves, we find that many objects in our sample show a  platform during the dim state,  lasting between $4\sim7$ years (e.g., see J101152.98+544206.4 as the best example). For the majority of our sample, we find consistent trends between the optical  and infrared light curves, suggesting that the CL phenomenon is caused by changes in accretion disk emission  rather than dust obscuration.   In addition, the turn-on and turn-off transition timescale is not significantly different,  and the observed timescale  is consistent with the cooling/heating front propagation timescale.  This evidence can be used to distinguish some models proposed to explain the CL phenomenon.  Our discovery significantly increases the previously identified recurrent CL AGN sample from eleven objects to nineteen, demonstrating that some AGNs can enter dormancy and reawaken on timescales of a few years, which provides useful information for understanding episodic AGN accretion. A uniform investigation of known CL AGNs is essential for fully understanding their long-term variations and minimizing selection bias.

%\begin{acknowledgments}
%We thank the anonymous referee for the helpful comments and suggestions.
This work is supported by the National Research Foundation of Korea (NRF) grant funded by the Korean government (MEST) (No. 2019R1A6A1A10073437), the Basic Science Research Program through the National Research Foundation of Korean Government (2021R1A2C3008486).  H.G. is supported by the National Key R\&D Program of China, No. 2022YFF0503402, and the Future Network Partner Program, CAS, No. 018GJHZ2022029FN, Overseas Center Platform Projects, CAS, No. 178GJHZ2023184MI.  J.J.  is supported by the National Natural Science Foundation of China 12403022.  M.K. is supported by Hebei Natural Science Foundation. A2024205020.

This work is based in part on observations of GN-2022A-FT-213 and GN-2022B-Q-305 obtained at the Gemini North Observatory, which is operated by the Association of Universities for Research in Astronomy (AURA) under a cooperative agreement with the NSF on behalf of the Gemini partnership: the National Science Foundation (United States), the National Research Council (Canada), CONICYT (Chile), the Australian Research Council (Australia), Ministerio da Ciencia e Tecnologia (Brazil) and Ministerio de Ciencia, Tecnologia e Innovacion Productiva (Argentina). 

This work utilizes the data from the ZTF. ZTF is supported by the National Science Foundation under Grant No. AST-2034437 and a collaboration including Caltech, IPAC, the Weizmann Institute for Science, the Oskar Klein Center at Stockholm University, the University of Maryland, Deutsches Elektronen-Synchrotron and Humboldt University, the TANGO Consortium of Taiwan, the University of Wisconsin at Milwaukee, Trinity College Dublin, Lawrence Livermore National Laboratories, and IN2P3, France. Operations are conducted by COO, IPAC, and UW. This research has made use of the NASA/IPAC Infrared Science Archive, which is funded by the National Aeronautics and Space Administration and operated by the California Institute of Technology. 

This work makes use of data products obtained from the Guoshoujing Telescope (LAMOST). LAMOST is a National Major Scientific Project built by the Chinese Academy of Sciences. Funding for the project has been provided by the National Development and Reform Commission. LAMOST is operated and managed by the National Astronomical Observatories, Chinese Academy of Sciences.

%\end{acknowledgments}

\facility{ZTF, Gemini, MDM, LAMOST} 

\software{Astropy \citep{Astropy13,Astropy18,Astropy22}, {\tt PyQSOFit} \citep{Guo18}, {\tt qso\_fit} \citep{Butler11}, {\tt PyCALI} \citep{Li14}.}

\appendix

\vspace{5mm}

\bibliography{ref.bib}\label{sec:ref}

\begin{thebibliography}{}
\expandafter\ifx\csname natexlab\endcsname\relax\def\natexlab#1{#1}\fi
\providecommand{\url}[1]{\href{#1}{#1}}
\providecommand{\dodoi}[1]{doi:~\href{http://doi.org/#1}{\nolinkurl{#1}}}
\providecommand{\doeprint}[1]{\href{http://ascl.net/#1}{\nolinkurl{http://ascl.net/#1}}}
\providecommand{\doarXiv}[1]{\href{https://arxiv.org/abs/#1}{\nolinkurl{https://arxiv.org/abs/#1}}}

\bibitem[{{Ai} {et~al.}(2016){Ai}, {Wu}, {Yang}, {Yang}, {Wang}, {Guo}, {Zuo},
  {Dong}, {Zhang}, {Yuan}, {Song}, {Wang}, {Dong}, {Yang}, {-Wu}, {Shen},
  {Shi}, {He}, {Lei}, {Li}, {Luo}, {Zhao}, \& {Zhang}}]{Ai16}
{Ai}, Y.~L., {Wu}, X.-B., {Yang}, J., {et~al.} 2016, \aj, 151, 24,
  \dodoi{10.3847/0004-6256/151/2/24}

\bibitem[{{Almeida} {et~al.}(2023){Almeida}, {Anderson},
  {Argudo-Fern{\'a}ndez}, {Badenes}, {Barger}, {Barrera-Ballesteros}, {Bender},
  {Benitez}, {Besser}, {Bird}, {Bizyaev}, {Blanton}, {Bochanski}, {Bovy},
  {Brandt}, {Brownstein}, {Buchner}, {Bulbul}, {Burchett}, {Cano D{\'\i}az},
  {Carlberg}, {Casey}, {Chandra}, {Cherinka}, {Chiappini}, {Coker}, {Comparat},
  {Conroy}, {Contardo}, {Cortes}, {Covey}, {Crane}, {Cunha}, {Dabbieri},
  {Davidson}, {Davis}, {de Andrade Queiroz}, {De Lee}, {M{\'e}ndez Delgado},
  {Demasi}, {Di Mille}, {Donor}, {Dow}, {Dwelly}, {Eracleous}, {Eriksen},
  {Fan}, {Farr}, {Frederick}, {Fries}, {Frinchaboy}, {G{\"a}nsicke}, {Ge},
  {Gonz{\'a}lez {\'A}vila}, {Grabowski}, {Grier}, {Guiglion}, {Gupta}, {Hall},
  {Hawkins}, {Hayes}, {Hermes}, {Hern{\'a}ndez-Garc{\'\i}a}, {Hogg},
  {Holtzman}, {Ibarra-Medel}, {Ji}, {Jofre}, {Johnson}, {Jones}, {Kinemuchi},
  {Kluge}, {Koekemoer}, {Kollmeier}, {Kounkel}, {Krishnarao}, {Krumpe},
  {Lacerna}, {Lago}, {Laporte}, {Liu}, {Liu}, {Liu}, {Lopes}, {Macktoobian},
  {Majewski}, {Malanushenko}, {Maoz}, {Masseron}, {Masters}, {Matijevic},
  {McBride}, {Medan}, {Merloni}, {Morrison}, {Myers}, {M{\'e}sz{\'a}ros},
  {Negrete}, {Nidever}, {Nitschelm}, {Oravetz}, {Oravetz}, {Pan}, {Peng},
  {Pinsonneault}, {Pogge}, {Qiu}, {Ramirez}, {Rix}, {Fern{\'a}ndez Rosso},
  {Runnoe}, {Salvato}, {Sanchez}, {Santana}, {Saydjari}, {Sayres},
  {Schlaufman}, {Schneider}, {Schwope}, {Serna}, {Shen}, {Sobeck}, {Song},
  {Souto}, {Spoo}, {Stassun}, {Steinmetz}, {Straumit}, {Stringfellow},
  {S{\'a}nchez-Gallego}, {Taghizadeh-Popp}, {Tayar}, {Thakar}, {Tissera},
  {Tkachenko}, {Hernandez Toledo}, {Trakhtenbrot}, {Fern{\'a}ndez-Trincado},
  {Troup}, {Trump}, {Tuttle}, {Ulloa}, {Vazquez-Mata}, {Vera Alfaro},
  {Villanova}, {Wachter}, {Weijmans}, {Wheeler}, {Wilson}, {Wojno}, {Wolf},
  {Xue}, {Ybarra}, {Zari}, \& {Zasowski}}]{Almeida23}
{Almeida}, A., {Anderson}, S.~F., {Argudo-Fern{\'a}ndez}, M., {et~al.} 2023,
  \apjs, 267, 44, \dodoi{10.3847/1538-4365/acda98}

\bibitem[{{Antonucci}(1993)}]{Antonucci93}
{Antonucci}, R. 1993, \araa, 31, 473,
  \dodoi{10.1146/annurev.aa.31.090193.002353}

\bibitem[{{Astropy Collaboration} {et~al.}(2013){Astropy Collaboration},
  {Robitaille}, {Tollerud}, {Greenfield}, {Droettboom}, {Bray}, {Aldcroft},
  {Davis}, {Ginsburg}, {Price-Whelan}, {Kerzendorf}, {Conley}, {Crighton},
  {Barbary}, {Muna}, {Ferguson}, {Grollier}, {Parikh}, {Nair}, {Unther},
  {Deil}, {Woillez}, {Conseil}, {Kramer}, {Turner}, {Singer}, {Fox}, {Weaver},
  {Zabalza}, {Edwards}, {Azalee Bostroem}, {Burke}, {Casey}, {Crawford},
  {Dencheva}, {Ely}, {Jenness}, {Labrie}, {Lim}, {Pierfederici}, {Pontzen},
  {Ptak}, {Refsdal}, {Servillat}, \& {Streicher}}]{Astropy13}
{Astropy Collaboration}, {Robitaille}, T.~P., {Tollerud}, E.~J., {et~al.} 2013,
  \aap, 558, A33, \dodoi{10.1051/0004-6361/201322068}

\bibitem[{{Astropy Collaboration} {et~al.}(2018){Astropy Collaboration},
  {Price-Whelan}, {Sip{\H{o}}cz}, {G{\"u}nther}, {Lim}, {Crawford}, {Conseil},
  {Shupe}, {Craig}, {Dencheva}, {Ginsburg}, {VanderPlas}, {Bradley},
  {P{\'e}rez-Su{\'a}rez}, {de Val-Borro}, {Aldcroft}, {Cruz}, {Robitaille},
  {Tollerud}, {Ardelean}, {Babej}, {Bach}, {Bachetti}, {Bakanov}, {Bamford},
  {Barentsen}, {Barmby}, {Baumbach}, {Berry}, {Biscani}, {Boquien}, {Bostroem},
  {Bouma}, {Brammer}, {Bray}, {Breytenbach}, {Buddelmeijer}, {Burke},
  {Calderone}, {Cano Rodr{\'\i}guez}, {Cara}, {Cardoso}, {Cheedella}, {Copin},
  {Corrales}, {Crichton}, {D'Avella}, {Deil}, {Depagne}, {Dietrich}, {Donath},
  {Droettboom}, {Earl}, {Erben}, {Fabbro}, {Ferreira}, {Finethy}, {Fox},
  {Garrison}, {Gibbons}, {Goldstein}, {Gommers}, {Greco}, {Greenfield},
  {Groener}, {Grollier}, {Hagen}, {Hirst}, {Homeier}, {Horton}, {Hosseinzadeh},
  {Hu}, {Hunkeler}, {Ivezi{\'c}}, {Jain}, {Jenness}, {Kanarek}, {Kendrew},
  {Kern}, {Kerzendorf}, {Khvalko}, {King}, {Kirkby}, {Kulkarni}, {Kumar},
  {Lee}, {Lenz}, {Littlefair}, {Ma}, {Macleod}, {Mastropietro}, {McCully},
  {Montagnac}, {Morris}, {Mueller}, {Mumford}, {Muna}, {Murphy}, {Nelson},
  {Nguyen}, {Ninan}, {N{\"o}the}, {Ogaz}, {Oh}, {Parejko}, {Parley}, {Pascual},
  {Patil}, {Patil}, {Plunkett}, {Prochaska}, {Rastogi}, {Reddy Janga},
  {Sabater}, {Sakurikar}, {Seifert}, {Sherbert}, {Sherwood-Taylor}, {Shih},
  {Sick}, {Silbiger}, {Singanamalla}, {Singer}, {Sladen}, {Sooley},
  {Sornarajah}, {Streicher}, {Teuben}, {Thomas}, {Tremblay}, {Turner},
  {Terr{\'o}n}, {van Kerkwijk}, {de la Vega}, {Watkins}, {Weaver}, {Whitmore},
  {Woillez}, {Zabalza}, \& {Astropy Contributors}}]{Astropy18}
{Astropy Collaboration}, {Price-Whelan}, A.~M., {Sip{\H{o}}cz}, B.~M., {et~al.}
  2018, \aj, 156, 123, \dodoi{10.3847/1538-3881/aabc4f}

\bibitem[{{Astropy Collaboration} {et~al.}(2022){Astropy Collaboration},
  {Price-Whelan}, {Lim}, {Earl}, {Starkman}, {Bradley}, {Shupe}, {Patil},
  {Corrales}, {Brasseur}, {N{\"o}the}, {Donath}, {Tollerud}, {Morris},
  {Ginsburg}, {Vaher}, {Weaver}, {Tocknell}, {Jamieson}, {van Kerkwijk},
  {Robitaille}, {Merry}, {Bachetti}, {G{\"u}nther}, {Aldcroft},
  {Alvarado-Montes}, {Archibald}, {B{\'o}di}, {Bapat}, {Barentsen},
  {Baz{\'a}n}, {Biswas}, {Boquien}, {Burke}, {Cara}, {Cara}, {Conroy},
  {Conseil}, {Craig}, {Cross}, {Cruz}, {D'Eugenio}, {Dencheva}, {Devillepoix},
  {Dietrich}, {Eigenbrot}, {Erben}, {Ferreira}, {Foreman-Mackey}, {Fox},
  {Freij}, {Garg}, {Geda}, {Glattly}, {Gondhalekar}, {Gordon}, {Grant},
  {Greenfield}, {Groener}, {Guest}, {Gurovich}, {Handberg}, {Hart},
  {Hatfield-Dodds}, {Homeier}, {Hosseinzadeh}, {Jenness}, {Jones}, {Joseph},
  {Kalmbach}, {Karamehmetoglu}, {Ka{\l}uszy{\'n}ski}, {Kelley}, {Kern},
  {Kerzendorf}, {Koch}, {Kulumani}, {Lee}, {Ly}, {Ma}, {MacBride}, {Maljaars},
  {Muna}, {Murphy}, {Norman}, {O'Steen}, {Oman}, {Pacifici}, {Pascual},
  {Pascual-Granado}, {Patil}, {Perren}, {Pickering}, {Rastogi}, {Roulston},
  {Ryan}, {Rykoff}, {Sabater}, {Sakurikar}, {Salgado}, {Sanghi}, {Saunders},
  {Savchenko}, {Schwardt}, {Seifert-Eckert}, {Shih}, {Jain}, {Shukla}, {Sick},
  {Simpson}, {Singanamalla}, {Singer}, {Singhal}, {Sinha}, {Sip{\H{o}}cz},
  {Spitler}, {Stansby}, {Streicher}, {{\v{S}}umak}, {Swinbank}, {Taranu},
  {Tewary}, {Tremblay}, {de Val-Borro}, {Van Kooten}, {Vasovi{\'c}}, {Verma},
  {de Miranda Cardoso}, {Williams}, {Wilson}, {Winkel}, {Wood-Vasey}, {Xue},
  {Yoachim}, {Zhang}, {Zonca}, \& {Astropy Project Contributors}}]{Astropy22}
{Astropy Collaboration}, {Price-Whelan}, A.~M., {Lim}, P.~L., {et~al.} 2022,
  \apj, 935, 167, \dodoi{10.3847/1538-4357/ac7c74}

\bibitem[{{Bellm} {et~al.}(2019){Bellm}, {Kulkarni}, {Graham}, {Dekany},
  {Smith}, {Riddle}, {Masci}, {Helou}, {Prince}, {Adams}, {Barbarino},
  {Barlow}, {Bauer}, {Beck}, {Belicki}, {Biswas}, {Blagorodnova}, {Bodewits},
  {Bolin}, {Brinnel}, {Brooke}, {Bue}, {Bulla}, {Burruss}, {Cenko}, {Chang},
  {Connolly}, {Coughlin}, {Cromer}, {Cunningham}, {De}, {Delacroix}, {Desai},
  {Duev}, {Eadie}, {Farnham}, {Feeney}, {Feindt}, {Flynn}, {Franckowiak},
  {Frederick}, {Fremling}, {Gal-Yam}, {Gezari}, {Giomi}, {Goldstein},
  {Golkhou}, {Goobar}, {Groom}, {Hacopians}, {Hale}, {Henning}, {Ho}, {Hover},
  {Howell}, {Hung}, {Huppenkothen}, {Imel}, {Ip}, {Ivezi{\'c}}, {Jackson},
  {Jones}, {Juric}, {Kasliwal}, {Kaspi}, {Kaye}, {Kelley}, {Kowalski},
  {Kramer}, {Kupfer}, {Landry}, {Laher}, {Lee}, {Lin}, {Lin}, {Lunnan},
  {Giomi}, {Mahabal}, {Mao}, {Miller}, {Monkewitz}, {Murphy}, {Ngeow},
  {Nordin}, {Nugent}, {Ofek}, {Patterson}, {Penprase}, {Porter}, {Rauch},
  {Rebbapragada}, {Reiley}, {Rigault}, {Rodriguez}, {van Roestel}, {Rusholme},
  {van Santen}, {Schulze}, {Shupe}, {Singer}, {Soumagnac}, {Stein}, {Surace},
  {Sollerman}, {Szkody}, {Taddia}, {Terek}, {Van Sistine}, {van Velzen},
  {Vestrand}, {Walters}, {Ward}, {Ye}, {Yu}, {Yan}, \& {Zolkower}}]{Bellm19}
{Bellm}, E.~C., {Kulkarni}, S.~R., {Graham}, M.~J., {et~al.} 2019, \pasp, 131,
  018002, \dodoi{10.1088/1538-3873/aaecbe}

\bibitem[{{Bennert} {et~al.}(2002){Bennert}, {Falcke}, {Schulz}, {Wilson}, \&
  {Wills}}]{Bennert02}
{Bennert}, N., {Falcke}, H., {Schulz}, H., {Wilson}, A.~S., \& {Wills}, B.~J.
  2002, \apjl, 574, L105, \dodoi{10.1086/342420}

\bibitem[{{Boroson} \& {Green}(1992)}]{Boroson92}
{Boroson}, T.~A., \& {Green}, R.~F. 1992, \apjs, 80, 109,
  \dodoi{10.1086/191661}

\bibitem[{{Butler} \& {Bloom}(2011)}]{Butler11}
{Butler}, N.~R., \& {Bloom}, J.~S. 2011, \aj, 141, 93,
  \dodoi{10.1088/0004-6256/141/3/93}

\bibitem[{{Denney} {et~al.}(2014){Denney}, {De Rosa}, {Croxall}, {Gupta},
  {Bentz}, {Fausnaugh}, {Grier}, {Martini}, {Mathur}, {Peterson}, {Pogge}, \&
  {Shappee}}]{Denney14}
{Denney}, K.~D., {De Rosa}, G., {Croxall}, K., {et~al.} 2014, \apj, 796, 134,
  \dodoi{10.1088/0004-637X/796/2/134}

\bibitem[{{Dexter} \& {Begelman}(2019)}]{Dexter19}
{Dexter}, J., \& {Begelman}, M.~C. 2019, \mnras, 483, L17,
  \dodoi{10.1093/mnrasl/sly213}

\bibitem[{{Dong} {et~al.}(2024){Dong}, {Zhang}, {Gu}, {Sun}, \&
  {Zheng}}]{Dong24}
{Dong}, Q., {Zhang}, Z.-X., {Gu}, W.-M., {Sun}, M., \& {Zheng}, Y.-G. 2024,
  arXiv e-prints, arXiv:2408.07335, \dodoi{10.48550/arXiv.2408.07335}

\bibitem[{{Dong} {et~al.}(2018){Dong}, {Wu}, {Ai}, {Yang}, {Yang}, {Wang},
  {Zhang}, {Luo}, {Xu}, {Yuan}, {Zhang}, {Wang}, {Wang}, {Li}, {Zuo}, {Hou},
  {Guo}, {Kong}, {Chen}, {Wu}, {Yang}, \& {Yang}}]{Dong18}
{Dong}, X.~Y., {Wu}, X.-B., {Ai}, Y.~L., {et~al.} 2018, \aj, 155, 189,
  \dodoi{10.3847/1538-3881/aab5ae}

\bibitem[{{Drake} {et~al.}(2009){Drake}, {Djorgovski}, {Mahabal}, {Beshore},
  {Larson}, {Graham}, {Williams}, {Christensen}, {Catelan}, {Boattini},
  {Gibbs}, {Hill}, \& {Kowalski}}]{Drake09}
{Drake}, A.~J., {Djorgovski}, S.~G., {Mahabal}, A., {et~al.} 2009, \apj, 696,
  870, \dodoi{10.1088/0004-637X/696/1/870}

\bibitem[{{Dubus} {et~al.}(2001){Dubus}, {Hameury}, \& {Lasota}}]{Dubus01}
{Dubus}, G., {Hameury}, J.~M., \& {Lasota}, J.~P. 2001, \aap, 373, 251,
  \dodoi{10.1051/0004-6361:20010632}

\bibitem[{{Feng J.} {et~al.}(2021){Feng J.}, {Cao}, {Li}, \& {Gu}}]{Feng21c}
{Feng J.}, J., {Cao}, X., {Li}, J.-w., \& {Gu}, W.-M. 2021, \apj, 916, 61,
  \dodoi{10.3847/1538-4357/ac07a6}

\bibitem[{{Graham} {et~al.}(2020){Graham}, {Ross}, {Stern}, {Drake},
  {McKernan}, {Ford}, {Djorgovski}, {Mahabal}, {Glikman}, {Larson}, \&
  {Christensen}}]{Graham20}
{Graham}, M.~J., {Ross}, N.~P., {Stern}, D., {et~al.} 2020, \mnras, 491, 4925,
  \dodoi{10.1093/mnras/stz3244}

\bibitem[{{Green} {et~al.}(2022){Green}, {Pulgarin-Duque}, {Anderson},
  {MacLeod}, {Eracleous}, {Ruan}, {Runnoe}, {Graham}, {Roulston}, {Schneider},
  {Ahlf}, {Bizyaev}, {Brownstein}, {del Casal}, {Dodd}, {Hoover}, {Matt},
  {Merloni}, {Pan}, {Ramirez}, {Ridder}, \& {Moseley}}]{Green22}
{Green}, P.~J., {Pulgarin-Duque}, L., {Anderson}, S.~F., {et~al.} 2022, \apj,
  933, 180, \dodoi{10.3847/1538-4357/ac743f}

\bibitem[{{Guo} {et~al.}(2018){Guo}, {Shen}, \& {Wang}}]{Guo18}
{Guo}, H., {Shen}, Y., \& {Wang}, S. 2018, {PyQSOFit: Python code to fit the
  spectrum of quasars}, Astrophysics Source Code Library, record ascl:1809.008.
\newblock \doeprint{1809.008}

\bibitem[{{Guo} {et~al.}(2019){Guo}, {Sun}, {Liu}, {Wang}, {Kong}, {Wang},
  {Sheng}, \& {He}}]{Guo19}
{Guo}, H., {Sun}, M., {Liu}, X., {et~al.} 2019, \apjl, 883, L44,
  \dodoi{10.3847/2041-8213/ab4138}

\bibitem[{{Guo} {et~al.}(2020{\natexlab{a}}){Guo}, {Peng}, {Zhang}, {Burke},
  {Liu}, {Sun}, {Wang}, {Kong}, {Sheng}, {Wang}, {He}, \& {Gu}}]{Guo20}
{Guo}, H., {Peng}, J., {Zhang}, K., {et~al.} 2020{\natexlab{a}}, \apj, 905, 52,
  \dodoi{10.3847/1538-4357/abc2ce}

\bibitem[{{Guo} {et~al.}(2020{\natexlab{b}}){Guo}, {Shen}, {He}, {Wang}, {Liu},
  {Wang}, {Sun}, {Yang}, {Kong}, \& {Sheng}}]{Guo20a}
{Guo}, H., {Shen}, Y., {He}, Z., {et~al.} 2020{\natexlab{b}}, \apj, 888, 58,
  \dodoi{10.3847/1538-4357/ab5db0}

\bibitem[{{Guo} {et~al.}(2024{\natexlab{a}}){Guo}, {Zou}, {Fawcett}, {Canning},
  {Juneau}, {Davis}, {Alexander}, {Jiang}, {Aguilar}, {Ahlen}, {Brooks},
  {Claybaugh}, {de la Macorra}, {Doel}, {Fanning}, {Forero-Romero}, {Gontcho A
  Gontcho}, {Honscheid}, {Kisner}, {Kremin}, {Landriau}, {Meisner}, {Miquel},
  {Moustakas}, {Nie}, {Pan}, {Poppett}, {Prada}, {Rezaie}, {Rossi}, {Siudek},
  {Sanchez}, {Schubnell}, {Seo}, {Sui}, {Tarl{\'e}}, \& {Zhou}}]{Guo24}
{Guo}, W.-J., {Zou}, H., {Fawcett}, V.~A., {et~al.} 2024{\natexlab{a}}, \apjs,
  270, 26, \dodoi{10.3847/1538-4365/ad118a}

\bibitem[{{Guo} {et~al.}(2024{\natexlab{b}}){Guo}, {Zou}, {Greenwell},
  {Alexander}, {Fawcett}, {Pan}, {Siudek}, {Aguilar}, {Ahlen}, {Brooks},
  {Claybaugh}, {Dawson}, {De La Macorra}, {Doel}, {Font-Ribera}, {Gaztanaga},
  {Gontcho}, {Gutierrez}, {Kehoe}, {Kisner}, {Landriau}, {Le Guillou},
  {Manera}, {Meisner}, {Mique}, {Moustakas}, {Prada}, {Rossi}, {Sanchez},
  {Schubnell}, {Sprayberry}, {Sui}, {Tarle}, {Weaver}, {Xiao}, \&
  {Zou}}]{Guo24b}
{Guo}, W.-J., {Zou}, H., {Greenwell}, C.~L., {et~al.} 2024{\natexlab{b}}, arXiv
  e-prints, arXiv:2408.00402, \dodoi{10.48550/arXiv.2408.00402}

\bibitem[{{Hon} {et~al.}(2022){Hon}, {Wolf}, {Onken}, {Webster}, \&
  {Auchettl}}]{Hon22}
{Hon}, W.~J., {Wolf}, C., {Onken}, C.~A., {Webster}, R., \& {Auchettl}, K.
  2022, \mnras, 511, 54, \dodoi{10.1093/mnras/stab3694}

\bibitem[{{Hutsem{\'e}kers} {et~al.}(2019){Hutsem{\'e}kers}, {Ag{\'\i}s
  Gonz{\'a}lez}, {Marin}, {Sluse}, {Ramos Almeida}, \& {Acosta
  Pulido}}]{Hutsemekers19}
{Hutsem{\'e}kers}, D., {Ag{\'\i}s Gonz{\'a}lez}, B., {Marin}, F., {et~al.}
  2019, \aap, 625, A54, \dodoi{10.1051/0004-6361/201834633}

\bibitem[{{Jiang} \& {Blaes}(2020)}]{Jiang20}
{Jiang}, Y.-F., \& {Blaes}, O. 2020, \apj, 900, 25,
  \dodoi{10.3847/1538-4357/aba4b7}

\bibitem[{{Jiang} {et~al.}(2019){Jiang}, {Blaes}, {Stone}, \&
  {Davis}}]{Jiang19}
{Jiang}, Y.-F., {Blaes}, O., {Stone}, J.~M., \& {Davis}, S.~W. 2019, \apj, 885,
  144, \dodoi{10.3847/1538-4357/ab4a00}

\bibitem[{{Jiang} {et~al.}(2016){Jiang}, {Davis}, \& {Stone}}]{Jiang16}
{Jiang}, Y.-F., {Davis}, S.~W., \& {Stone}, J.~M. 2016, \apj, 827, 10,
  \dodoi{10.3847/0004-637X/827/1/10}

\bibitem[{{Jin} {et~al.}(2023){Jin}, {Wu}, {Fu}, {Yao}, {Ai}, {Feng}, {He},
  {Ma}, {Pang}, {Zhu}, {Zhang}, {Yuan}, \& {Huo}}]{Jin23}
{Jin}, J.-J., {Wu}, X.-B., {Fu}, Y., {et~al.} 2023, \apjs, 265, 25,
  \dodoi{10.3847/1538-4365/acaf89}

\bibitem[{{Kochanek} {et~al.}(2017){Kochanek}, {Shappee}, {Stanek}, {Holoien},
  {Thompson}, {Prieto}, {Dong}, {Shields}, {Will}, {Britt}, {Perzanowski}, \&
  {Pojma{\'n}ski}}]{Kochanek17}
{Kochanek}, C.~S., {Shappee}, B.~J., {Stanek}, K.~Z., {et~al.} 2017, \pasp,
  129, 104502, \dodoi{10.1088/1538-3873/aa80d9}

\bibitem[{{Krolik}(1999)}]{Krolik99}
{Krolik}, J.~H. 1999, {Active galactic nuclei : from the central black hole to
  the galactic environment}

\bibitem[{{LaMassa} {et~al.}(2015){LaMassa}, {Cales}, {Moran}, {Myers},
  {Richards}, {Eracleous}, {Heckman}, {Gallo}, \& {Urry}}]{laMassa15}
{LaMassa}, S.~M., {Cales}, S., {Moran}, E.~C., {et~al.} 2015, \apj, 800, 144,
  \dodoi{10.1088/0004-637X/800/2/144}

\bibitem[{{Lawrence}(1991)}]{Lawrence91}
{Lawrence}, A. 1991, \mnras, 252, 586, \dodoi{10.1093/mnras/252.4.586}

\bibitem[{{Li} \& {Cao}(2019)}]{Li-Cao19}
{Li}, J., \& {Cao}, X. 2019, \apj, 872, 149, \dodoi{10.3847/1538-4357/ab0207}

\bibitem[{{Li} {et~al.}(2014){Li}, {Wang}, {Hu}, {Du}, \& {Bai}}]{Li14}
{Li}, Y.-R., {Wang}, J.-M., {Hu}, C., {Du}, P., \& {Bai}, J.-M. 2014, \apjl,
  786, L6, \dodoi{10.1088/2041-8205/786/1/L6}

\bibitem[{Lintott {et~al.}(2009)Lintott, Schawinski, Keel, Van~Arkel, Bennert,
  Edmondson, Thomas, Smith, Herbert, Jarvis, Virani, Andreescu, Bamford, Land,
  Murray, Nichol, Raddick, Slosar, Szalay, \& Vandenberg}]{Lintott09}
Lintott, C.~J., Schawinski, K., Keel, W., {et~al.} 2009, Monthly Notices of the
  Royal Astronomical Society, 399, 129,
  \dodoi{10.1111/j.1365-2966.2009.15299.x}

\bibitem[{{L{\'o}pez-Navas} {et~al.}(2022){L{\'o}pez-Navas},
  {Mart{\'\i}nez-Aldama}, {Bernal}, {S{\'a}nchez-S{\'a}ez}, {Ar{\'e}valo},
  {Graham}, {Hern{\'a}ndez-Garc{\'\i}a}, {Lira}, \& {Rojas Lobos}}]{Navas22}
{L{\'o}pez-Navas}, E., {Mart{\'\i}nez-Aldama}, M.~L., {Bernal}, S., {et~al.}
  2022, \mnras, 513, L57, \dodoi{10.1093/mnrasl/slac033}

\bibitem[{{L{\'o}pez-Navas} {et~al.}(2023){L{\'o}pez-Navas},
  {S{\'a}nchez-S{\'a}ez}, {Ar{\'e}valo}, {Bernal}, {Graham},
  {Hern{\'a}ndez-Garc{\'\i}a}, {Homan}, {Krumpe}, {Lamer}, {Lira},
  {Mart{\'\i}nez-Aldama}, {Merloni}, {R{\'\i}os}, {Salvato}, {Stern}, \&
  {Tub{\'\i}n-Arenas}}]{Navas23b}
{L{\'o}pez-Navas}, E., {S{\'a}nchez-S{\'a}ez}, P., {Ar{\'e}valo}, P., {et~al.}
  2023, \mnras, 524, 188, \dodoi{10.1093/mnras/stad1893}

\bibitem[{{Luo} {et~al.}(2015){Luo}, {Zhao}, {Zhao}, {Deng}, {Liu}, {Jing},
  {Wang}, {Zhang}, {Shi}, {Cui}, {Chu}, {Li}, {Bai}, {Wu}, {Cai}, {Cao}, {Cao},
  {Carlin}, {Chen}, {Chen}, {Chen}, {Chen}, {Chen}, {Chen}, {Chen},
  {Christlieb}, {Chu}, {Cui}, {Dong}, {Du}, {Fan}, {Feng}, {Fu}, {Gao}, {Gong},
  {Gu}, {Guo}, {Han}, {He}, {Hou}, {Hou}, {Hou}, {Hu}, {Hu}, {Hu}, {Huo},
  {Jia}, {Jiang}, {Jiang}, {Jiang}, {Jin}, {Kong}, {Kong}, {Lei}, {Li}, {Li},
  {Li}, {Li}, {Li}, {Li}, {Li}, {Li}, {Li}, {Li}, {Li}, {Li}, {Liang}, {Lin},
  {Liu}, {Liu}, {Liu}, {Liu}, {Lu}, {Luo}, {Mao}, {Newberg}, {Ni}, {Qi}, {Qi},
  {Shen}, {Shi}, {Song}, {Song}, {Su}, {Su}, {Tang}, {Tao}, {Tian}, {Wang},
  {Wang}, {Wang}, {Wang}, {Wang}, {Wang}, {Wang}, {Wang}, {Wang}, {Wang},
  {Wang}, {Wang}, {Wang}, {Wang}, {Wang}, {Wang}, {Wang}, {Wang}, {Wang},
  {Wang}, {Wei}, {Wei}, {Wu}, {Wu}, {Wu}, {Wu}, {Xing}, {Xu}, {Xu}, {Xu},
  {Yan}, {Yang}, {Yang}, {Yang}, {Yang}, {Yao}, {Yu}, {Yuan}, {Yuan}, {Yuan},
  {Yuan}, {Zhai}, {Zhang}, {Zhang}, {Zhang}, {Zhang}, {Zhang}, {Zhang},
  {Zhang}, {Zhang}, {Zhao}, {Zhou}, {Zhou}, {Zhu}, {Zhu}, {Zou}, \&
  {Zuo}}]{Luo15}
{Luo}, A.~L., {Zhao}, Y.-H., {Zhao}, G., {et~al.} 2015, Research in Astronomy
  and Astrophysics, 15, 1095, \dodoi{10.1088/1674-4527/15/8/002}

\bibitem[{{MacLeod} {et~al.}(2016){MacLeod}, {Ross}, {Lawrence}, {Goad},
  {Horne}, {Burgett}, {Chambers}, {Flewelling}, {Hodapp}, {Kaiser}, {Magnier},
  {Wainscoat}, \& {Waters}}]{Macleod16}
{MacLeod}, C.~L., {Ross}, N.~P., {Lawrence}, A., {et~al.} 2016, \mnras, 457,
  389, \dodoi{10.1093/mnras/stv2997}

\bibitem[{{MacLeod} {et~al.}(2019){MacLeod}, {Green}, {Anderson}, {Bruce},
  {Eracleous}, {Graham}, {Homan}, {Lawrence}, {LeBleu}, {Ross}, {Ruan},
  {Runnoe}, {Stern}, {Burgett}, {Chambers}, {Kaiser}, {Magnier}, \&
  {Metcalfe}}]{Macleod19}
{MacLeod}, C.~L., {Green}, P.~J., {Anderson}, S.~F., {et~al.} 2019, \apj, 874,
  8, \dodoi{10.3847/1538-4357/ab05e2}

\bibitem[{{Mathur} {et~al.}(2018){Mathur}, {Denney}, {Gupta}, {Vestergaard},
  {De Rosa}, {Krongold}, {Nicastro}, {Collinson}, {Goad}, {Korista}, {Pogge},
  \& {Peterson}}]{Mathur18}
{Mathur}, S., {Denney}, K.~D., {Gupta}, A., {et~al.} 2018, \apj, 866, 123,
  \dodoi{10.3847/1538-4357/aadd91}

\bibitem[{{Noda} \& {Done}(2018)}]{Noda18}
{Noda}, H., \& {Done}, C. 2018, \mnras, 480, 3898,
  \dodoi{10.1093/mnras/sty2032}

\bibitem[{{Oh} {et~al.}(2015){Oh}, {Yi}, {Schawinski}, {Koss}, {Trakhtenbrot},
  \& {Soto}}]{Oh15}
{Oh}, K., {Yi}, S.~K., {Schawinski}, K., {et~al.} 2015, \apjs, 219, 1,
  \dodoi{10.1088/0067-0049/219/1/1}

\bibitem[{{Osterbrock}(1977)}]{Osterbrock77}
{Osterbrock}, D.~E. 1977, \apj, 215, 733, \dodoi{10.1086/155407}

\bibitem[{{Pan} {et~al.}(2021){Pan}, {Li}, \& {Cao}}]{Pan21}
{Pan}, X., {Li}, S.-L., \& {Cao}, X. 2021, \apj, 910, 97,
  \dodoi{10.3847/1538-4357/abe766}

\bibitem[{{Raimundo} {et~al.}(2019){Raimundo}, {Vestergaard}, {Koay},
  {Lawther}, {Casasola}, \& {Peterson}}]{Raimundo19}
{Raimundo}, S.~I., {Vestergaard}, M., {Koay}, J.~Y., {et~al.} 2019, \mnras,
  486, 123, \dodoi{10.1093/mnras/stz852}

\bibitem[{{Ren} {et~al.}(2024){Ren}, {Guo}, {Shen}, {Silverman}, {Burke},
  {Wang}, \& {Wang}}]{Ren24}
{Ren}, W., {Guo}, H., {Shen}, Y., {et~al.} 2024, arXiv e-prints,
  arXiv:2406.17598, \dodoi{10.48550/arXiv.2406.17598}

\bibitem[{{Ricci} \& {Trakhtenbrot}(2022)}]{Ricci22}
{Ricci}, C., \& {Trakhtenbrot}, B. 2022, arXiv e-prints, arXiv:2211.05132,
  \dodoi{10.48550/arXiv.2211.05132}

\bibitem[{{Ricci} {et~al.}(2022){Ricci}, {Ananna}, {Temple}, {Urry}, {Koss},
  {Trakhtenbrot}, {Ueda}, {Stern}, {Bauer}, {Treister}, {Privon}, {Oh},
  {Paltani}, {Stalevski}, {Ho}, {Fabian}, {Mushotzky}, {Chang}, {Ricci},
  {Kakkad}, {Sartori}, {Baer}, {Caglar}, {Powell}, \& {Harrison}}]{Ricci22b}
{Ricci}, C., {Ananna}, T.~T., {Temple}, M.~J., {et~al.} 2022, \apj, 938, 67,
  \dodoi{10.3847/1538-4357/ac8e67}

\bibitem[{{Roig} {et~al.}(2014){Roig}, {Blanton}, \& {Ross}}]{Roig14}
{Roig}, B., {Blanton}, M.~R., \& {Ross}, N.~P. 2014, \apj, 781, 72,
  \dodoi{10.1088/0004-637X/781/2/72}

\bibitem[{{Ross} {et~al.}(2018){Ross}, {Ford}, {Graham}, {McKernan}, {Stern},
  {Meisner}, {Assef}, {Dey}, {Drake}, {Jun}, \& {Lang}}]{Ross18}
{Ross}, N.~P., {Ford}, K.~E.~S., {Graham}, M., {et~al.} 2018, \mnras, 480,
  4468, \dodoi{10.1093/mnras/sty2002}

\bibitem[{{Ruan} {et~al.}(2019){Ruan}, {Anderson}, {Eracleous}, {Green},
  {Haggard}, {MacLeod}, {Runnoe}, \& {Sobolewska}}]{Ruan19}
{Ruan}, J.~J., {Anderson}, S.~F., {Eracleous}, M., {et~al.} 2019, \apj, 883,
  76, \dodoi{10.3847/1538-4357/ab3c1a}

\bibitem[{{Rumbaugh} {et~al.}(2018){Rumbaugh}, {Shen}, {Morganson}, {Liu},
  {Banerji}, {McMahon}, {Abdalla}, {Benoit-L{\'e}vy}, {Bertin}, {Brooks},
  {Buckley-Geer}, {Capozzi}, {Carnero Rosell}, {Carrasco Kind}, {Carretero},
  {Cunha}, {D'Andrea}, {da Costa}, {DePoy}, {Desai}, {Doel}, {Frieman},
  {Garc{\'\i}a-Bellido}, {Gruen}, {Gruendl}, {Gschwend}, {Gutierrez},
  {Honscheid}, {James}, {Kuehn}, {Kuhlmann}, {Kuropatkin}, {Lima}, {Maia},
  {Marshall}, {Martini}, {Menanteau}, {Plazas}, {Reil}, {Roodman}, {Sanchez},
  {Scarpine}, {Schindler}, {Schubnell}, {Sheldon}, {Smith}, {Soares-Santos},
  {Sobreira}, {Suchyta}, {Swanson}, {Walker}, {Wester}, \& {DES
  Collaboration}}]{Rumbaugh18}
{Rumbaugh}, N., {Shen}, Y., {Morganson}, E., {et~al.} 2018, \apj, 854, 160,
  \dodoi{10.3847/1538-4357/aaa9b6}

\bibitem[{{Runco} {et~al.}(2016){Runco}, {Cosens}, {Bennert}, {Scott},
  {Komossa}, {Malkan}, {Lazarova}, {Auger}, {Treu}, \& {Park}}]{Runco16}
{Runco}, J.~N., {Cosens}, M., {Bennert}, V.~N., {et~al.} 2016, \apj, 821, 33,
  \dodoi{10.3847/0004-637X/821/1/33}

\bibitem[{{Runnoe} {et~al.}(2016){Runnoe}, {Cales}, {Ruan}, {Eracleous},
  {Anderson}, {Shen}, {Green}, {Morganson}, {LaMassa}, {Greene}, {Dwelly},
  {Schneider}, {Merloni}, {Georgakakis}, \& {Roman-Lopes}}]{Runnoe16}
{Runnoe}, J.~C., {Cales}, S., {Ruan}, J.~J., {et~al.} 2016, \mnras, 455, 1691,
  \dodoi{10.1093/mnras/stv2385}

\bibitem[{{S{\'a}nchez-S{\'a}ez} {et~al.}(2021){S{\'a}nchez-S{\'a}ez}, {Lira},
  {Mart{\'\i}}, {S{\'a}nchez-Pi}, {Arredondo}, {Bauer}, {Bayo},
  {Cabrera-Vives}, {Donoso-Oliva}, {Est{\'e}vez}, {Eyheramendy}, {F{\"o}rster},
  {Hern{\'a}ndez-Garc{\'\i}a}, {Arancibia}, {P{\'e}rez-Carrasco},
  {Sep{\'u}lveda}, \& {Vergara}}]{Sanchez21}
{S{\'a}nchez-S{\'a}ez}, P., {Lira}, H., {Mart{\'\i}}, L., {et~al.} 2021, \aj,
  162, 206, \dodoi{10.3847/1538-3881/ac1426}

\bibitem[{{Shakura} \& {Sunyaev}(1973)}]{Shakura73}
{Shakura}, N.~I., \& {Sunyaev}, R.~A. 1973, \aap, 24, 337

\bibitem[{{Shen}(2021)}]{Shen21}
{Shen}, Y. 2021, \apj, 921, 70, \dodoi{10.3847/1538-4357/ac1ce4}

\bibitem[{{Shen} \& {Liu}(2012)}]{Shen12}
{Shen}, Y., \& {Liu}, X. 2012, \apj, 753, 125,
  \dodoi{10.1088/0004-637X/753/2/125}

\bibitem[{{Shen} {et~al.}(2011){Shen}, {Richards}, {Strauss}, {Hall},
  {Schneider}, {Snedden}, {Bizyaev}, {Brewington}, {Malanushenko},
  {Malanushenko}, {Oravetz}, {Pan}, \& {Simmons}}]{Shen11}
{Shen}, Y., {Richards}, G.~T., {Strauss}, M.~A., {et~al.} 2011, \apjs, 194, 45,
  \dodoi{10.1088/0067-0049/194/2/45}

\bibitem[{{Shen} {et~al.}(2019){Shen}, {Hall}, {Horne}, {Zhu}, {McGreer},
  {Simm}, {Trump}, {Kinemuchi}, {Brandt}, {Green}, {Grier}, {Guo}, {Ho},
  {Homayouni}, {Jiang}, {I-Hsiu Li}, {Morganson}, {Petitjean}, {Richards},
  {Schneider}, {Starkey}, {Wang}, {Chambers}, {Kaiser}, {Kudritzki}, {Magnier},
  \& {Waters}}]{Shen19}
{Shen}, Y., {Hall}, P.~B., {Horne}, K., {et~al.} 2019, \apjs, 241, 34,
  \dodoi{10.3847/1538-4365/ab074f}

\bibitem[{{Sheng} {et~al.}(2017){Sheng}, {Wang}, {Jiang}, {Yang}, {Yan}, {Dou},
  \& {Peng}}]{Sheng17}
{Sheng}, Z., {Wang}, T., {Jiang}, N., {et~al.} 2017, \apjl, 846, L7,
  \dodoi{10.3847/2041-8213/aa85de}

\bibitem[{{Sniegowska} {et~al.}(2020){Sniegowska}, {Czerny}, {Bon}, \&
  {Bon}}]{Sniegowska20}
{Sniegowska}, M., {Czerny}, B., {Bon}, E., \& {Bon}, N. 2020, \aap, 641, A167,
  \dodoi{10.1051/0004-6361/202038575}

\bibitem[{{Stern} {et~al.}(2018){Stern}, {McKernan}, {Graham}, {Ford}, {Ross},
  {Meisner}, {Assef}, {Balokovi{\'c}}, {Brightman}, {Dey}, {Drake},
  {Djorgovski}, {Eisenhardt}, \& {Jun}}]{Stern18}
{Stern}, D., {McKernan}, B., {Graham}, M.~J., {et~al.} 2018, \apj, 864, 27,
  \dodoi{10.3847/1538-4357/aac726}

\bibitem[{{Temple} {et~al.}(2023){Temple}, {Ricci}, {Koss}, {Trakhtenbrot},
  {Bauer}, {Mushotzky}, {Rojas}, {Caglar}, {Harrison}, {Oh}, {Padilla
  Gonzalez}, {Powell}, {Ricci}, {Riffel}, {Stern}, \& {Urry}}]{Temple23}
{Temple}, M.~J., {Ricci}, C., {Koss}, M.~J., {et~al.} 2023, \mnras, 518, 2938,
  \dodoi{10.1093/mnras/stac3279}

\bibitem[{{Tonry} {et~al.}(2018){Tonry}, {Denneau}, {Heinze}, {Stalder},
  {Smith}, {Smartt}, {Stubbs}, {Weiland}, \& {Rest}}]{Tonry18}
{Tonry}, J.~L., {Denneau}, L., {Heinze}, A.~N., {et~al.} 2018, \pasp, 130,
  064505, \dodoi{10.1088/1538-3873/aabadf}

\bibitem[{{Veronese} {et~al.}(2024){Veronese}, {Vignali}, {Severgnini},
  {Matzeu}, \& {Cignoni}}]{Veronese24}
{Veronese}, S., {Vignali}, C., {Severgnini}, P., {Matzeu}, G.~A., \& {Cignoni},
  M. 2024, \aap, 683, A131, \dodoi{10.1051/0004-6361/202348098}

\bibitem[{{Wang} {et~al.}(2020){Wang}, {Shen}, {Jiang}, {Grier}, {Horne},
  {Homayouni}, {Peterson}, {Trump}, {Brandt}, {Hall}, {Ho}, {Li}, {Hernandez
  Santisteban}, {Kinemuchi}, {McGreer}, \& {Schneider}}]{Wang20}
{Wang}, S., {Shen}, Y., {Jiang}, L., {et~al.} 2020, \apj, 903, 51,
  \dodoi{10.3847/1538-4357/abb36d}

\bibitem[{{Wang} {et~al.}(2024){Wang}, {Woo}, {Gallo}, {Guo}, {Son}, {Kong},
  {Mandal}, {Cho}, {Kim}, \& {Shin}}]{Wang24}
{Wang}, S., {Woo}, J.-H., {Gallo}, E., {et~al.} 2024, \apj, 966, 128,
  \dodoi{10.3847/1538-4357/ad3049}

\bibitem[{{Wang-J} {et~al.}(2024){Wang-J}, {Xu}, {Cao}, {Gao}, {Xie}, \&
  {Wei}}]{Wang-J24}
{Wang-J}, J., {Xu}, D.~W., {Cao}, X., {et~al.} 2024, arXiv e-prints,
  arXiv:2405.10663, \dodoi{10.48550/arXiv.2405.10663}

\bibitem[{{Yan} {et~al.}(2019){Yan}, {Wang}, {Jiang}, {Stern}, {Dou},
  {Fremling}, {Graham}, {Drake}, {Yang}, {Burdge}, \& {Kasliwal}}]{Yan19}
{Yan}, L., {Wang}, T., {Jiang}, N., {et~al.} 2019, \apj, 874, 44,
  \dodoi{10.3847/1538-4357/ab074b}

\bibitem[{{Yang} {et~al.}(2024){Yang}, {Green}, {Wu}, {Eracleous}, {Jiang}, \&
  {Fu}}]{Yang24}
{Yang}, Q., {Green}, P.~J., {Wu}, X.-B., {et~al.} 2024, arXiv e-prints,
  arXiv:2408.16183, \dodoi{10.48550/arXiv.2408.16183}

\bibitem[{{Yang} {et~al.}(2018){Yang}, {Wu}, {Fan}, {Jiang}, {McGreer},
  {Shangguan}, {Yao}, {Wang}, {Joshi}, {Green}, {Wang}, {Feng}, {Fu}, {Yang},
  \& {Liu}}]{Yang18}
{Yang}, Q., {Wu}, X.-B., {Fan}, X., {et~al.} 2018, \apj, 862, 109,
  \dodoi{10.3847/1538-4357/aaca3a}

\bibitem[{{Yang} {et~al.}(2020){Yang}, {Shen}, {Liu}, {Aguena}, {Annis},
  {Avila}, {Banerji}, {Bertin}, {Brooks}, {Burke}, {Carnero Rosell}, {Carrasco
  Kind}, {da Costa}, {De Vicente}, {Desai}, {Diehl}, {Doel}, {Flaugher},
  {Fosalba}, {Frieman}, {Garcia-Bellido}, {Gerdes}, {Gruen}, {Gruendl},
  {Gschwend}, {Gutierrez}, {Hinton}, {Hollowood}, {Honscheid}, {Kuropatkin},
  {Maia}, {March}, {Marshall}, {Martini}, {Melchior}, {Menanteau}, {Miquel},
  {Paz-Chinchon}, {Malag{\'o}n}, {Romer}, {Sanchez}, {Scarpine}, {Schubnell},
  {Serrano}, {Sevilla}, {Smith}, {Suchyta}, {Tarle}, {Varga}, \&
  {Wilkinson}}]{Yang20b}
{Yang}, Q., {Shen}, Y., {Liu}, X., {et~al.} 2020, \apj, 900, 58,
  \dodoi{10.3847/1538-4357/aba59b}

\bibitem[{{Yang} {et~al.}(2023){Yang}, {Green}, {MacLeod}, {Plotkin},
  {Anderson}, {Bieryla}, {Civano}, {Eracleous}, {Graham}, {Ruan}, {Runnoe}, \&
  {Zhao}}]{Yang23}
{Yang}, Q., {Green}, P.~J., {MacLeod}, C.~L., {et~al.} 2023, \apj, 953, 61,
  \dodoi{10.3847/1538-4357/acdedd}

\bibitem[{{Yao} {et~al.}(2019){Yao}, {Wu}, {Ai}, {Yang}, {Yang}, {Dong},
  {Joshi}, {Wang}, {Feng}, {Fu}, {Hou}, {Luo}, {Kong}, {Liu}, {Zhao}, {Zhang},
  {Yuan}, \& {Shen}}]{Yao19}
{Yao}, S., {Wu}, X.-B., {Ai}, Y.~L., {et~al.} 2019, \apjs, 240, 6,
  \dodoi{10.3847/1538-4365/aaef88}

\bibitem[{{Yip} {et~al.}(2004{\natexlab{a}}){Yip}, {Connolly}, {Szalay},
  {Budav{\'a}ri}, {SubbaRao}, {Frieman}, {Nichol}, {Hopkins}, {York},
  {Okamura}, {Brinkmann}, {Csabai}, {Thakar}, {Fukugita}, \&
  {Ivezi{\'c}}}]{Yip04a}
{Yip}, C.~W., {Connolly}, A.~J., {Szalay}, A.~S., {et~al.} 2004{\natexlab{a}},
  \aj, 128, 585, \dodoi{10.1086/422429}

\bibitem[{{Yip} {et~al.}(2004{\natexlab{b}}){Yip}, {Connolly}, {Vanden Berk},
  {Ma}, {Frieman}, {SubbaRao}, {Szalay}, {Richards}, {Hall}, {Schneider},
  {Hopkins}, {Trump}, \& {Brinkmann}}]{Yip04b}
{Yip}, C.~W., {Connolly}, A.~J., {Vanden Berk}, D.~E., {et~al.}
  2004{\natexlab{b}}, \aj, 128, 2603, \dodoi{10.1086/425626}

\bibitem[{{Zeltyn} {et~al.}(2022){Zeltyn}, {Trakhtenbrot}, {Eracleous},
  {Runnoe}, {Trump}, {Stern}, {Shen}, {Hern{\'a}ndez-Garc{\'\i}a}, {Bauer},
  {Yang}, {Dwelly}, {Ricci}, {Green}, {Anderson}, {Assef}, {Guolo}, {MacLeod},
  {Davis}, {Fries}, {Gezari}, {Grogin}, {Homan}, {Koekemoer}, {Krumpe},
  {LaMassa}, {Liu}, {Merloni}, {Mart{\'\i}nez-Aldama}, {Schneider}, {Temple},
  {Brownstein}, {Ibarra-Medel}, {Burke}, {Pellegrino}, \&
  {Kollmeier}}]{Zeltyn22}
{Zeltyn}, G., {Trakhtenbrot}, B., {Eracleous}, M., {et~al.} 2022, \apjl, 939,
  L16, \dodoi{10.3847/2041-8213/ac9a47}

\bibitem[{{Zeltyn} {et~al.}(2024){Zeltyn}, {Trakhtenbrot}, {Eracleous}, {Yang},
  {Green}, {Anderson}, {LaMassa}, {Runnoe}, {Assef}, {Bauer}, {Brandt},
  {Davis}, {Frederick}, {Fries}, {Graham}, {Grogin}, {Guolo},
  {Hern{\'a}ndez-Garc{\'\i}a}, {Koekemoer}, {Krumpe}, {Liu},
  {Mart{\'\i}nez-Aldama}, {Ricci}, {Schneider}, {Shen}, {{\'S}niegowska},
  {Temple}, {Trump}, {Xue}, {Brownstein}, {Dwelly}, {Morrison}, {Bizyaev},
  {Pan}, \& {Kollmeier}}]{Zeltyn24}
---. 2024, \apj, 966, 85, \dodoi{10.3847/1538-4357/ad2f30}

\bibitem[{{Zhao} {et~al.}(2012){Zhao}, {Zhao}, {Chu}, {Jing}, \&
  {Deng}}]{Zhao12}
{Zhao}, G., {Zhao}, Y.-H., {Chu}, Y.-Q., {Jing}, Y.-P., \& {Deng}, L.-C. 2012,
  Research in Astronomy and Astrophysics, 12, 723,
  \dodoi{10.1088/1674-4527/12/7/002}

\end{thebibliography}
\bibliographystyle{aasjournal}

\end{document}